\documentclass[aps,onecolumn,nofootinbib,floatfix]{revtex4-2}

\usepackage{graphicx}
\usepackage{xcolor}
\usepackage{url}
\usepackage{amsmath}
\usepackage{enumitem}
\usepackage{comment}
\usepackage[section]{placeins}  

\usepackage[symbol]{footmisc}

\usepackage{bbm}
\usepackage{cancel}
\usepackage[utf8]{inputenc}
\usepackage[english]{babel}
\usepackage{tikz} 
\usepackage{dcolumn}  
\usepackage{mathtools} 
\usepackage{empheq}
\usepackage{bm}        
\usepackage{amssymb}    
\usepackage{bbold}
\usepackage{relsize}
\usepackage[T1]{fontenc}
\usepackage{mathtools}
\usepackage{amsfonts,amssymb,amsthm}
 
\usepackage{multirow}
\usepackage{float} 
\usepackage{xcolor}
\usepackage{tikz}

\long\def\comment#1{}

\def\cite#1{\citep{#1}}

\newcommand{\be}{\begin{equation}}
\newcommand{\ee}{\end{equation}}
\newcommand{\bea}{\begin{eqnarray}}
\newcommand{\eea}{\end{eqnarray}}

\usepackage{ulem}

\newcommand{\Arrow}[1]{%
\parbox{#1}{\tikz{\draw[->](0,0)--(#1,0);}}
}

\begin{document}

\begin{abstract}
 Experiments on decision making under uncertainty are known to display a classical pattern of risk aversion and risk seeking  referred to as "fourfold pattern" (or "reflection effect") , but recent experiments varying the speed and order of mental processing  have brought to light a more nuanced phenomenology. We model experiments though a Bayesian formalization of the anchor-and-adjust heuristic observed in empirical studies on cognitive bias. Using only elementary assumptions on constrained information processing, we are able to infer three separate effects found in recent observations: (1) the reported enhancement of the fourfold pattern for quicker decision processes; (2) the observed decrease of fluctuations  for  slower 
decision-making trials; (3) the reported dependence of the outcome on the order in which options are processed.  The application of Bayesian modeling offers a solution to recent empirical riddles by bridging two heretofore separate domains of experimental inquiry on  bounded rationality.
\end{abstract}

\author{Francesco Fumarola\footnote{Equal contributors \label{footnote 1}}}
\affiliation{
%Laboratory for Neural Computation and Adaptation, 
RIKEN Center for Brain Science, Wako, Saitama 351-0198, Japan}
\author{Lukasz Kusmierz$^{\ref{footnote 1}}$}
\affiliation{Allen Institute, Seattle,
%615 Westlake Ave N, Seattle,
WA 98109, United States}
\author{Ronald B. Dekker}
\affiliation{Department of Experimental Psychology, University of Oxford, OX2 6BW Oxford, United Kingdom}

\title{"Bayesian anchoring" and the fourfold pattern of risk attitudes}

\date{\today}
\pacs{}
\maketitle

\vspace{20pt}

\section{Introduction}

\begin{figure}
\centering
\includegraphics[width=.4\linewidth]{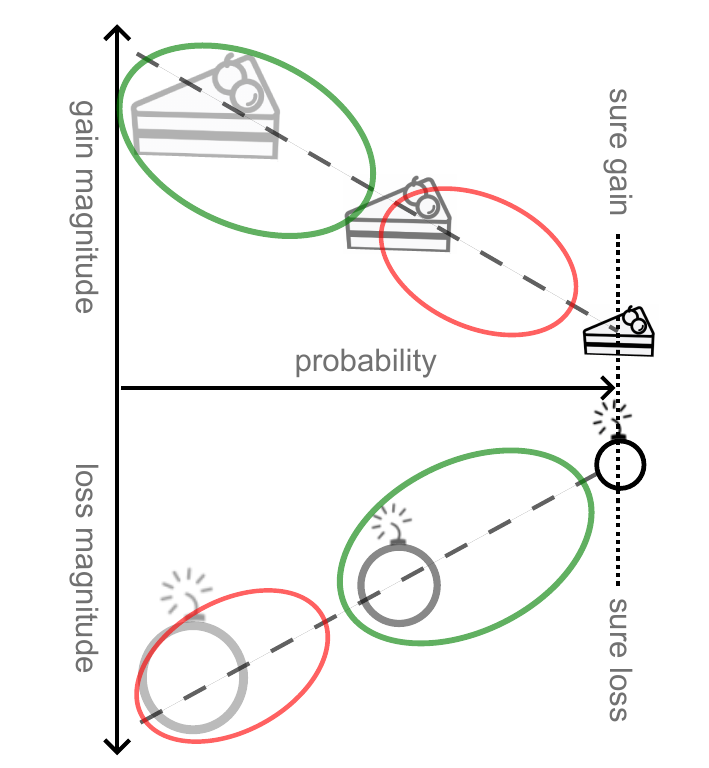}
\caption{
Schematic depiction of the fourfold pattern of risk attitudes. Dashed lines outline risky choices with equal expected values; along such a line, the green ellipse highlights the preferred region and the red ellipse the disfavored region, for gains and losses separately. Although a cake and a bomb will be used to picturize positive and negative outcome, the common choice in experiments is for loss and gain to be monetary, easing the comparison of expected values among  options. In  gain experiments, subjects favor very unlikely but very large gain (risk seeking) or sure outcome; in loss experiments, subjects favor likely but small losses (risk aversion).   
}
\label{fig:reflection}
\end{figure}

One of the foundational results on decision making under uncertainty  is the fourfold pattern of risk attitudes often referred to as reflection effect~\cite{kahneman2011thinking,barberis2013thirty}. 
 Given a choice between a risky and a risk-free  choice with identical average outcomes, subjects are risk-seeking when the possible gain is large (and unlikely) and when the loss is likely (but relatively small);  subjects are risk-averse when the possible gain is likely but relatively small, and when the loss is unlikely and large (this is summarized in Fig.~\ref{fig:reflection}). The neural mechanisms underlying the effect have long been an object of debate among neuroscientists~\cite{trepel2005prospect,fox2009prospect}. The effect is customarily described through phenomenological models, the most relevant being  Prospect Theory~\cite{kahneman1979prospect}, which  posits that subjects compute expected utilities  by weighing probabilities through a  nonlinear weight function. The function is unknown \textit{a priori} and must be specified as a parameter of the theory, though only its convexity/concavity  properties matter. Since the appearance of the theory, the literature has seen the proposal of both  variations~\cite{wakker1993axiomatization} and alternatives~\cite{chan2021decision}.

One of the main recent developments in research on human decision making is the emergence of a body of empirical  evidence that Prospect Theory alone is not able to account for. Such evidence was obtained by psychologists by  modifying details of the classic experimental procedure -- most notably  the  time allowed for decision-making and the order in which the options are presented to subjects. 

The former direction of inquiry concerns the influence of time constraints on individual preferences and choices (for an overview,~\cite{spiliopoulos2018bcd}). By randomly allocating subjects into time pressure conditions (i.e. varying randomly the completion deadline for tasks), the literature has identified the effects of time constraints on risky, social, and strategic behavior in such contexts as bargaining~\cite{sutter2003bargaining}, beauty contests~\cite{kocher2006time},  decisions under ambiguity~\cite{baillon2013ambiguity}, imitation in strategic games~\cite{buckert2017imitation}, and  bidding in auctions~\cite{el2019time}. 

What may be called "slow" risk tasks --  risky decisions made with no hindrance to tackling the full computation workload -- have been carefully isolated in experiments such as those of~\cite{kachelmeier1992examining},  which involve substantial monetary rewards for completing the tasks correctly and allow the full necessary time for decision making. These studies have confirmed the risk behavior predicted by Prospect Theory, showing that such behavior arises as a result of underlying prioritization, not short-term cognitive-load issues. It has long remained inconclusive, however, whether  constraints on information processing combine synergically with the fourfold pattern or affect the decision outcome in more complex ways~\cite{kirchler2017effect}. 

Recently, "fast" risk-taking  decisions have become  the object of a dedicated set of experimental tests by Kirchler et al.~\cite{kirchler2017effect}, who brought the random time approach to the analysis of risk behavior. The major findings reported are that 
(1) time pressure increases risk aversion for gains and risk taking for losses compared to time delay;  (2) fluctuations in the outcome are suppressed by granting subjects a longer time to decide.

One framework that should be mentioned here as being applied to multiple problems in psychology is the broad dichotomy between two alternative thinking processes: "System 1" (fast, intuitive, and emotional) and "System 2" (slower,  deliberative,  more  logical)~\cite{kahneman2011thinking}. The System 1 vs System 2 duality  offers a convenient language to broadly discuss a variety of different experiments.  For instance, one may say that the role of time pressure~\cite{kirchler2017effect}, increased cognitive load~\cite{otto2013curse}, or acute stress~\cite{otto2013working} during decision making is to increase the role of System 1, as System 2 would be computationally and energetically more expensive. 

 One would assume that fast decisions rely on a different psychological attitude than slower decisions, and are computed therefore on the basis of a different set of underlying risk perceptions. However, even in the absence of a dual process faster decisions would be affected by computational limitations and by the reduced ability to cope with cognitive load, leading to distinct outcomes. It is unclear at present how much, among the features of fast decision making, follows simply from  information-processing constraints. The  fact that performance on the Cognitive Reflection Test (CRT) and  variable decision time correlate identically with the observed extent of cognitive bias is intriguing but does not allow to draw  definite conclusions~\cite{frederick2005cognitive,kirchler2017effect}. 

The finding that  time pressure increases risk aversion for gains and risk taking for losses  can be phrased as stating that time constraints increase the fourfold pattern predicted by Prospect Theory, but the reasons for this trend remain open to multiple interpretations. Kirchler et al. interpret the finding  as confirming that System 1 gives rise to the fourfold pattern, but the specific features of System 1 that would do so are currently unknown.

Asides from experiments on time pressure, another direction of recent experimental inquiry concerns the effect of variations in the processing order of the options. Various  features of visual processing have long been known to  affect choice behavior~\cite{Armel2008, Gidlof2013, Krajbich2010, Shimojo2003} but one crucial  factor appears  to be the  sequence in which information is presented; this has  been shown to have an impact  in real-life setting such as personality judgment~\cite{Asch1946, Jones1968}, clinical decision making~\cite{Chapman1996, Cunnington1997}, and judicial decision making~\cite{KerstholtJackson1998}).

Most recently, \cite{kwak2018order} replicated the classical decision-making experiment we have described and additionally assessed the processing order of the options. They did so in two different ways: through eye-tracking, and by presenting the options sequentially. Both approaches led to the finding that the magnitude of loss/gain asymmetry effects is larger when the risk-free option is processed first. A mechanistic explanation for this asymmetry is  lacking, but a description in terms of System 1 vs System 2 dichotomy may be attempted. In particular,  \cite{kwak2018order} invokes  "changes in arousal and/or emotion induced by the initial perception of the frame information, i.e. to framing effects capable of modulating between  avoidance behaviors and approach behaviors.

In this paper, we point out a candidate for a common mechanism behind all of the above findings. The mechanism we propose is not novel -- it is the anchor-and-adjust heuristic~\cite{lieder2018anchoring} widely documented by a separate lineage of  experimental literature (reviewed further below). Formalizing that heuristic into what we term Bayesian Anchoring,  we calculate the outcome of time-constrained information processing in risky decision making and from that premise alone we derive the findings of ~\cite{kirchler2017effect}. 

The outline  of the  paper is as follows:

\begin{itemize}
\item 
 We sketch what is arguably the most basic Bayesian framework for time-pressured cognition (Sec.~\ref{sec:bayesian}), with a view to pursuing it to its furthest logical conclusions.  We feed into that framework (Sec. \ref{sec:application}) a cognitive prior taken from empirical observations (reviewed in Sec.~\ref{sec:anchoring}), thus setting the structure of Bayesian Anchoring. 

\item We reproduce the fourfold pattern of risk attitudes from Bayesian Anchoring (Sec.~\ref{sec:risky_decision}), providing a coherent explanation  to experimental findings on risky decision making under time pressure~\cite{kirchler2017effect}. We show that multi-step iterations gradually diminish the fourfold pattern, which explains the recent, unexplained findings of Kirchler et al. (Sec.~\ref{sec:multistep}). 
 
\item We then move to discuss what Ba ayesian anchoring theory would look like with priors different from the one we selected initially (Sec.~\ref{sec:general}), and  we additionally present a number of theorems concerning the set of Bayesian Anchoring priors that would yield predictions compatible with experiments.  

\item We finally summarize our contribution, compare Bayesian Anchoring to other modeling efforts in the literature, and discuss the new research directions it opens.  

\end{itemize}

\section{Modeling fast decisions}
\label{sec:bayesian}

The problem we focus on is simply stated. The subject  needs to make an informed guess on some not-immediately-known variable $S$ ("decision variable") that quantifies the stochastic gain or loss (utility) associated to a decision problem. This guess will informed by some set of relevant knowledge $K$, which may be thought of as the result of reviewing past experience and  simulating random outcomes accordingly. 

This  conceptual setup resembles for example that of Lieder et al.~\cite{lieder2018anchoring}, who proceed to develop a model based on the familiar theory of Markov Chain Monte Carlo (MCMC). However, here our focus is on choices constrained by limits on computation time, which makes a MCMC model inappropriate. Indeed, we cannot assume that adjustments are implemented through steps of the Metropolis-Hasting algorithm, which would require access to (possibly unnormalized) conditional density $\mathbf{P}(S|K)$ and hence access to all relevant knowledge at once. It would take substantial time and computational effort to access, analyze, and synthesize all relevant knowledge. Instead, we assume that in time-constrained decisions, adjustments are driven by 
an iterative process wherein the belief over $S$ is updated by integrating at every step a single piece of information from the set of relevant knowledge. We start by focusing on a tightly time-constrained regime, analyzing therefore a single such adjustment step. This may be thought of as a single round of simulations of the random outcome, used to update the utility based on the posterior from the realizations~\cite{einhorn1986decision}.  
 
Utility is most often represented in economics as taking  values on the real line. We thus  assume the random variable $S$ to be realized by real values $s$. In the very first approximation, every piece of information $K$  can be distilled into a pair of numbers $(x, \sigma)$ describing  the mean and standard deviation of possible values of $s$ that this piece information supports ($\sigma$ being  the measurement error). The corresponding generative model, or likelihood function, can be written as 

\be\rho( (x,\sigma)| s ) = g_{\sigma}(x - s)\ee

where $g_{\sigma}(z) = [\sqrt{2 \pi} \sigma ]^{-1} \exp(-z^2/(2\sigma^2))$ 
is a normal distribution with standard deviation $\sigma$. The 
assumption that the noise in the estimate of the value of interest 
is Gaussian is non-essential to results and is used here for mathematical 
convenience. 

The corresponding posterior probability density of $s$ can be expressed as

\be
\label{Bayes_rule}
\rho(s|(x,\sigma)) =
\frac{\rho((x,\sigma)|s) \rho_0(s) }{\rho(x,\sigma)}
=\frac{g_{\sigma}(x - s) \rho_0(s)}{\int ds g_{\sigma}(x - s)\rho_0(s)}
\ee

where we are abusing notation by denoting both likelihood and posterior  densities as $\rho$, with arguments and conditionalities to distinguish the  distributions, while we call $ \rho_0$ the distribution of prior beliefs. The next step consists in determining an ansatz on the function $ \rho_0(s)$ enforced by existing experimental knowledge.

To this aim, we will briefly review the relevant  phenomenology.

\section{Review of anchoring and adjustment}
\label{sec:anchoring}

The anchor-and-adjust heuristic is a simple heuristic widely observed in human cognition~\cite{lieder2018anchoring}. It is  usually 
associated with the purported irrationality 
of human behavior as it leads to cognitive biases that do not 
seem to follow rules of logic and probability as well as 
give rise to sub-optimal decision-making. 
While the concept has wide applicability, the case most relevant to this work is the one where  subjects, faced with uncertainty, start their estimation process 
from a very rough guess termed  \textit{anchor}, 
possibly suggested by the experimenter,
and then integrate additional information to 
adjust the estimate. 
Since often these adjustments are not sufficient, 
this leads to an observable bias that can be manipulated 
by the experimenter. 

For example, in one well-known experiment~\cite{tversky1974judgment} participants (high-school 
students) 
were divided into two groups and were asked to 
multiply either
\be
    1\times2\times3\times4\times5\times6\times7\times8
\ee
or
\be 8\times7\times6\times5\times4\times3\times2\times1
\ee 
The trick was that the participants were given only 
5 seconds to report the answer---way below the time 
required for most people to perform such mental calculations. 
Thus, they had to resort to heuristics. 
One may expect that people would perform a few steps 
of computation, starting from numbers on the left, 
and the extrapolate to estimate the full product. 
Since the adjustments are often insufficient, we should 
expect students to underestimate these products. 
Moreover, the first group should get lower values of the 
anchor and thus underestimate more.
Indeed, the median estimate for the first group was $512$, compared to $2250$ for the other group. 
This is obviously much lower than the correct result ($40320$).

\section{Anchoring as Bayesian inference with a spike-and-slab prior}
\label{sec:application}

In this section we formalize the Bayesian Anchoring model, describe its single-step realization, and 
discuss possible multi-step extensions.
\subsection{General considerations}
The phenomenology we just summarized can be formalized into a choice for  the cognitive prior figuring in Eq.~(\ref{Bayes_rule}). Given an initial guess (anchor) $s_0$, we posit that the 
belief over $X$ is stored as a (prior) density 

\be
\label{priorslab}
\rho_0(s) = p_0 \delta(s-s_0) + (1-p_0) g_{\gamma_0}(s-s_0) 
\ee

We will refer to this as spike-and-slab prior~\cite{mitchell1988bayesian,ishwaran2011consistency}, the "slab" component being here not uniform but modulated (for simplicity, as a Gaussian with variance $\gamma_0^2$). 

This prior can be interpreted as a mixture of two hypotheses 
about $S$. Under the first hypothesis ($\mathcal{H}_0$, with prior probability $P(\mathcal{H}_0)=p_0$)  the initial guess is correct, i.e., $\rho(s|\mathcal{H}_0) =  \delta(s-s_0)$. As a consequence, observing any information 
from the set of accessible knowledge cannot influence the 
initial guess, since we have 

\be \rho(s|(x,\sigma), \mathcal{H}_0) = \frac{g_{\sigma}(x - s) \rho(s|\mathcal{H}_0)}{\int d s g_{\sigma}(x - s)\rho(s|\mathcal{H}_0)}
=
\delta(s-s_0) 
=
\rho(s|\mathcal{H}_0)
\ee

In contrast to $\mathcal{H}_0$, the other hypothesis 
presumes non-zero variance of the initial guess $\rho(s|\mathcal{H}_1) = g_{\gamma_0}(s-s_0)$. 
Accordingly, $\mathcal{H}_1$ acknowledges the possibility that 
relevant knowledge can change the belief over $S$, since 
in general we have 

\be \rho(s|(x,\sigma),\mathcal{H}_1) \neq \rho(s|\mathcal{H}_1) \ee

The posterior distribution of $s$ has a form similar to the prior:

\be
\label{rho_s_1}
\rho(s|(x,\sigma)) = p_1 \delta(s-s_0) + (1-p_1) g_{\gamma_1}(s-s_1)
\ee

where $p_1$, $\gamma_1$, and $s_1$ are updated values 
of parameters that depend on the parameters 
of the prior as well as on the observation. In particular, $p_1=P(\mathcal{H}_0|(x,\sigma))$ is the 
posterior probability of the hypothesis $\mathcal{H}_0$:

\be 
p_1 = \frac{1}{1 + K^{-1}}
\ee 
where 
\be 
K = \frac{p_0}{1-p_0} \frac{g_{\sigma}(x-s_0)}{g_{\tilde{\sigma}}(x-s_0)}
\ee 
and 
\be 
\tilde{\sigma}^{2} = \sigma^{2} + \gamma_0^2
\ee 
The width of the slab 
decreases according to the formula
\be
\gamma_1^{-2} = \gamma_0^{-2} + \sigma^{-2}
\ee
The location of the slab shifts towards the observation as
\be 
s_1 = 
\frac{\gamma_0^{-2}}{\gamma_0^{-2} + \sigma^{-2}} s_0 
+
\frac{\sigma^{-2}}{\gamma_0^{-2} + \sigma^{-2}} x
\ee 
Since the location of the spike is constant, in general the location of the spike and Gaussian slab in the posterior do not coincide ($s_1 \neq s_0$). 

\subsection{Single adjustment step}

Let us first assume that just a single adjustment step has been made--a likely scenario when faced with very limited time. 
After the adjustment step the posterior has been calculated, 
but what value should be used to compare options and take a decision? 
Let us denote the point estimator used for decision-making to compare options as $\hat{s}$. 
If the observation confirms the anchor ($x=s_0$), the optimal estimator should remain the anchor, so we must have $ \hat{s}(0) = s_0$. On the other hand, any difference between $\hat{s}$ and $x$ must clearly come from a bias in the prior distribution $\rho_0(s)$, and in the limit where the prior is very flat ($p_0 \rightarrow 0$), we must have the asymptotic behavior $ \hat{s} \approx x$. 

These requirements are satisfied by a wide range of estimators and many 
results presented in this work should extend to other such estimators. For concreteness, in the following we choose the minimum mean squared error estimator, which amounts to calculating the expectation $\hat{s}=\langle s|x \rangle$ over the posterior distribution. Measuring the argument of the estimator by its displacement from the anchor, we define $ \hat{s}(x-s_0 ; s_0)  : = \langle s |x \rangle$, and will omit  explicit mention of the anchor as the second argument of $\hat{s}$ where obvious. 
The explicit expression for $\hat{s}$ reads
\be
\hat{s} = s_0 + \frac{1}{1+K} \frac{\gamma_0^2}{\sigma^2 + \gamma_0^2}(x-s_0).
\ee
Note that, due to the dependence on the Bayes factor $K$, 
$\hat{s}$ is a nonlinear function of $x-s_0$.  
This nonlinearity will be essential when we discuss 
the role of anchoring-with-adjustment in the 
fourfold pattern of risk attitudes 
(Section \ref{sec:risky_decision}).

\subsection{Multiple adjustment steps}

There are multiple ways in which one could extend our model to incorporate 
multiple adjustments steps. We need to specify an update rule and a stopping rule. 
The update rule describes how the new prior for the next step is formed from the 
previous prior given the observation. The most natural update rule seems to be to 
perform the exact Bayesian inference and to use the posterior as the new prior.
However, to perform Bayesian inference with multiple steps of 
adjustments the agent should track values of $p_n$, 
$\gamma_n$, $s_n$, 
and to remember the initial value of the anchor $s_0$. 
Additionally, in general the Gaussian prior of $\mathcal{H}_1$ 
is not a conjugate prior if the likelihood takes a different form. 
Thus, the inference may become more complex and 
may require approximations or sampling.

Another possibility is to perform a simple average over $\{x_i\}_{i=1..T}$ (where $T$ denotes the number of pseudo-observations made in order to perform adjustments) 
and then use this sample mean in the single adjustment step, utilizing anchoring slab-and-spike priors as described above. 

In the following we explore another hypothesis that humans rely on a simplified recursive inference strategy. 
First note that the process of multi-step anchoring-and-adjustment must at some point be terminated. 
The following stochastic rule could utilize the probabilistic 
form of the posterior in order to make a decision whether to continue or stop the 
process. 
A single sample is drawn, taken from the posterior distribution over 
hypotheses (i.e. $\mathcal{H}_0$ or $\mathcal{H}_1$). 
If $\mathcal{H}_0$ is chosen the adjustment process is terminated 
and the value $\hat{s}$ is reported.
Otherwise, another adjustment step is taken with a new prior that 
takes the form

\be 
\rho_1(s) = p_1 \delta(s-s_1) + (1-p_1) g_{\gamma_1}(s-s_1).
\ee

This formula introduces an important modification compared to the 
optimal Bayesian update. 
Namely, the position of the spike under $\mathcal{H}_0$ is 
updated to reflect the new most likely position of $s$. 
Such reanchoring is not essential for the consistency of Bayesian Anchoring theory but it simplifies the calculations and may help account for the fact that 
the initial anchor is rarely reported, even if it biases the final report. 
We hypothesize that the reanchoring position corresponds to the mean of the posterior slab ($s_1$), but other reanchoring methods can be formulated. We performed numerical simulations to check how expected bias and error change as functions of the number of sampling and adjustment steps, depending on the reanchoring method (Fig.~\ref{fig:multistep_simulations}).
Our simulation results indicate that reanchoring to the posterior slab mean quickly removes the bias. In contrast, reanchoring to the overall posterior mean ($\hat{s}$) leads to an extremely persistent bias.

Additionally, it is possible that instead of updating $p_0$ to $p_1$ according to the 
rules of Bayesian inference, $p_n$ is modulated by 
the internally or externally generated signal indicating 
a sense of urgency. 
Intuitively, the more urgent it is to finish the inference, the higher 
should be the value of $p_n$. This in turn makes it more likely to stop 
the adjustment process in the following step.

\begin{figure}
\centering
\includegraphics[width=.5\linewidth]{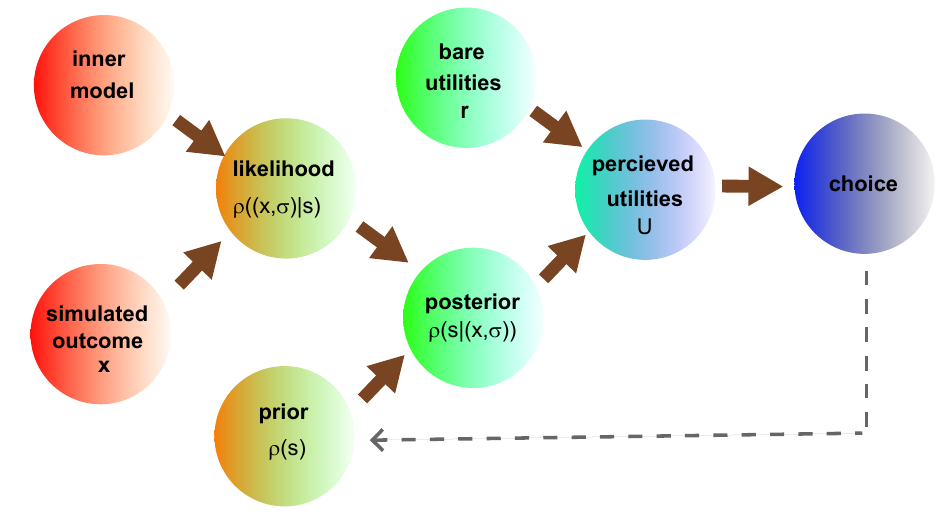}
\caption{Structure of cognitively biased decision making based on a subjective update of utilities. The update may be thought of as the result of reviewing past experience and  simulating random outcomes accordingly. The perceived utilities resulting from the process drive subjective choices. Sections~\ref{sec:bayesian}-\ref{sec:multistep}  of this article follow the flow of the model along the brown arrows in the drawing, from  assumptions about the internal variables all the way to inferred behavior. In Sec.~\ref{sec:general}, we show for completeness the results of proceeding the opposite way, starting from the observed behavior and working backwards to infer prior believes (gray dashed arrow).}
\label{fig:flow}
\end{figure}
 
\section{Emergence of the fourfold pattern of risk attitudes}
\label{sec:risky_decision}

\vspace{10pt}

We now apply the framework we developed to the  experimental setting routinely used to verify fourfold pattern of risk attitudes~\cite{kahneman1979prospect,kahneman2011thinking,barberis2013thirty}, where risky choices are associated to a binary outcome as shown in Table~\ref{table}. The utility acquired though a risky choice stochastically turns out to be  zero or to have a positive/negative value. We  will call $p$ the probability of the latter outcome.

\begin{table}[h!]
  \centering
\begin{tabular}{c  c  | c |  c | c }
  &  & \ Gain/loss value $x$  \ & \ Probability 
\ & \  $ \langle x \rangle_{\text{outcomes}}$   
\\ &&&&\\
  \hline  &&&& \\
   \multirow{2}*{Option A (risky) $ \longrightarrow$}  &  \ \ Outcome $1$ \ 
   & $r/p$ & $p$ &   \multirow{2}*{$r$}\\
~  &   \  \ Outcome 2  \ & $0$ & $1-p$ & ~ \\
  &&&& \\ \hline & & && \\
    Option B (risk-free) & \Arrow{1cm} & $r$ & $1$&
$r$ \end{tabular}
    \caption{Structure of a choice options in traditional experiments on economic decision~\cite{kahneman1979prospect}. Subjects are given two  options to choose from, a risky and a risk-free one. The last columns shows  the mean gain/loss, i.e. the average of the gain/loss value over objective outcomes, which is the same for both options.}
    \label{table}
\end{table}

As shown in Table \ref{table} subjects who choose the risk-free outcome receive or lose a value $x=r$ with probability $1$; subjects who choose the risky outcome  receive or lose value  $x = r/p$ with probability $p$ and $x = 0$ with probability $1-p$. 

To yield the perceived utility associated to a choice option, the point estimate $\hat{s}$  must be averaged over all outcomes of that option. The perceived utility of a given option is therefore

\be 
\label{U_general}
U^{\text{(option)}}  =   \langle \hat{s}(x - s_0, s_0) \rangle_{\text{outcomes}}
\ee

Although the anchor is usually not reported, it is natural then to expect that the gain/loss value stated  as the sure outcome of the riskless choice would be the anchor $s_0 = r$.  
Replacing this into Eq. \ref{U_general} for the two options $A$ and $B$ in Table \ref{table},

\bea 
\label{utility_main_text} 
U^{(A)} (p,r) &=&   \langle \hat{s}(x - s_0, s_0) \rangle_A = \langle \hat{s}(x - \langle x \rangle, \langle x \rangle) \rangle_A = p \ \hat{s}(r/p -r, r) + (1-p) \hat{s}(-r, r) 
\\ 
U^{(B)} (r) &=&  \langle \hat{s}(x - s_0, s_0) \rangle_B = \langle \hat{s}(x - \langle x \rangle, \langle x \rangle) \rangle_B =\hat{s}(0, r) = r
\eea

where we note that $U^{(B)}(r) = U^{(A)}(1,r)$ because setting the gain/loss probability to unity removes all risk from the risky option. 

 A decision is made by comparing the perceived utility of Eqs. \ref{utility_main_text} over all options available (Fig.~\ref{fig:flow}), that is, the experimentally observed decision depends on the sign of $U = U^{(A)} - U^{(B)}$, which is just $U^{A}$ measured with its $p=1$ value for a  a baseline.

Hence, the fourfold pattern of risk attitudes corresponds to the finding that the relative perceived utility $U$ flips sign as a function of the gain/loss probability $p$, becoming positive only for sufficiently large $p$ if $r$ is negative, and becoming  negative only for sufficiently large $p$ if $r$ is positive. 
Subjects will reject the risk-free option in favor of an uncertain gain only if that gain is sufficiently large, and will reject it in favor of a purely potential loss (which may ultimately not take place) only if that loss is sufficiently small (see Sec.~\ref{sec:perceived_utility} for details).

\begin{figure}
    \centering
    \includegraphics[width=.95\linewidth]{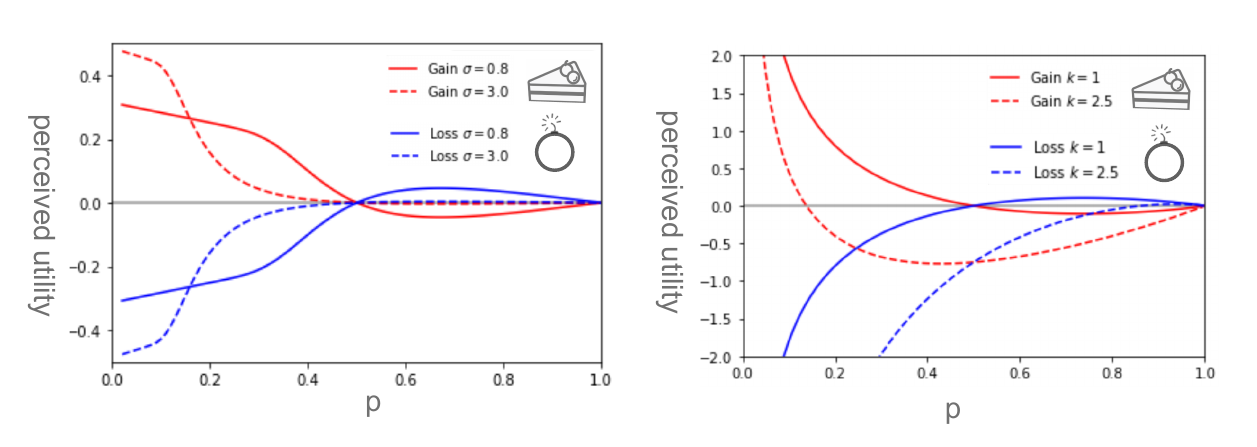}
    \caption{Perceived utility of a risky option with loss/gain probability $p$, measured relative to the riskless option with equivalent expected outcome. 
    \textit{Left:} Predictions of the anchoring theory. The main features of experiments, such as the transition between risk-aversion and risk-seeking and the associated reflection principle, are reproduced. \textit{Right:} Predictions of the  phenomenological model of~\cite{chan2021decision}  shown for comparison.  
    One difference between the two predictions is that the Bayesian model does not lead to $U \to \pm \infty$ with $p\to 0$. In its basic version, the Bayesian  model can only reproduce the symmetric version with $k=1$ and the bias towards risk aversion must be introduced ad hoc, similarly to the  model of~\cite{chan2021decision}. The parameter values used here are $p_0 = 0.2$ and $\gamma_0 = 10$.
    }
    \label{fig:perceived_utility}
\end{figure}

Simply plugging (\ref{priorslab}) into (\ref{Bayes_rule}) and the result into (\ref{utility_main_text}) immediately yields the comparative perceived utility plotted in Fig.~\ref{fig:perceived_utility}, left-hand panel (see Sec.~\ref{sec:spike_slab} for details). The utility indeed flips sign as expected from experiments, yielding risky choices in the gain domain only for high gains, and risky choices in the loss domain only for small losses. Thus, Bayesian Anchoring theory qualitatively reproduces the fourfold pattern 
of risk attitudes with an arguably simpler set of assumptions compared to 
Prospect Theory.

The main assumption has been the choice of the risk-free reward as an anchor, which  is  motivated by the certainty of the reward within the risk-free option. In other words, the reward associated to a risk-free option is assumed to be more "salient".  Within that assumption, the experimental results of~\cite{kwak2018order} are also arguably explained as a corollary.

Indeed, suppose that the risky option is presented first. Only after the risk-free option is also given, its saliency is recognized and the anchor set (or reset). This takes time and makes the process more liable to randomization. By contrast, if the first option is the risk-free one, the anchor is set before the second option is given, and the decision making is performed immediately upon hearing the second option. An alternative way to think about this is that two different aspects contribute to the determination of the anchor, saliency and primacy. The interference between these two effects is constructive if the risk-free option is presented first, destructive otherwise.  This agrees with the result of  experiments  where the processing order is monitored as in~\cite{kwak2018order}. 

Experiments that vary the processing speed have a more elaborate output and will be discussed next.  
 
\section{Time-dependent fluctuations: experiments and theory } 
\label{sec:multistep}

Kirchler et al.~\cite{kirchler2017effect} conducted a comprehensive study of the fourfold pattern of risk attitudes over a large sample, with real financial rewards and across multiple geographical areas.

One of their findings is that fluctuations in the response are enhanced by time pressure. In section 5 of~\cite{kirchler2017effect}, the different components of such fluctuations are teased out by breaking apart the effect of time pressure on risk preferences and the effect of time pressure on noisiness. 

Experiment includes a risky option and a non-risky option. The amount suggested for the non-risky option is given and requires no calculation, so in the Bayesian Anchoring framework it would correspond to the anchor $s_0$. 
Thus, $s_0$ is induced externally and as such will not cause much fluctuations. 

\begin{figure}
\centering
\includegraphics[width=.9\linewidth]{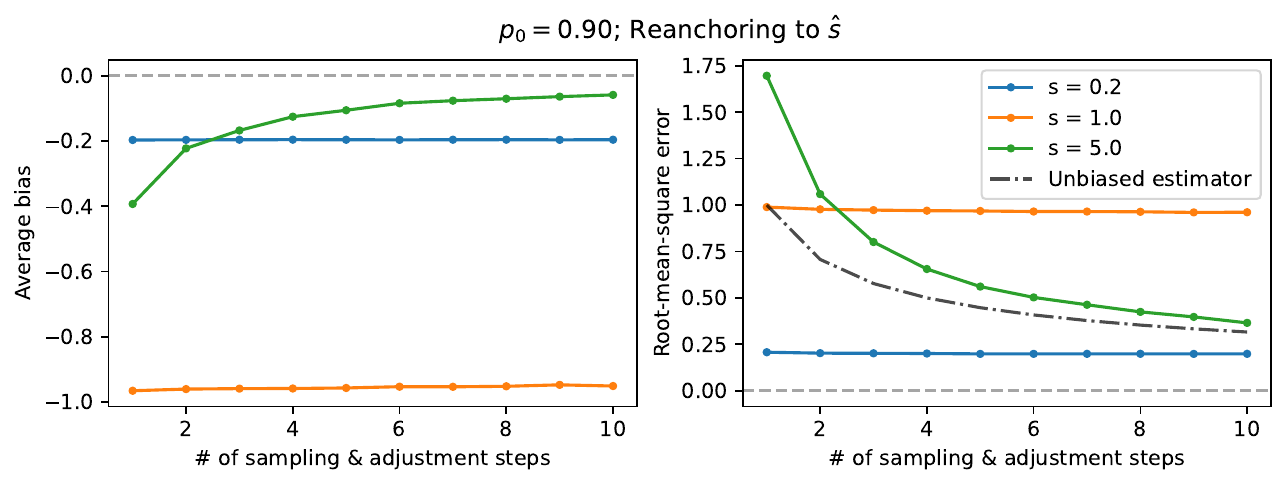}
\includegraphics[width=.9\linewidth]{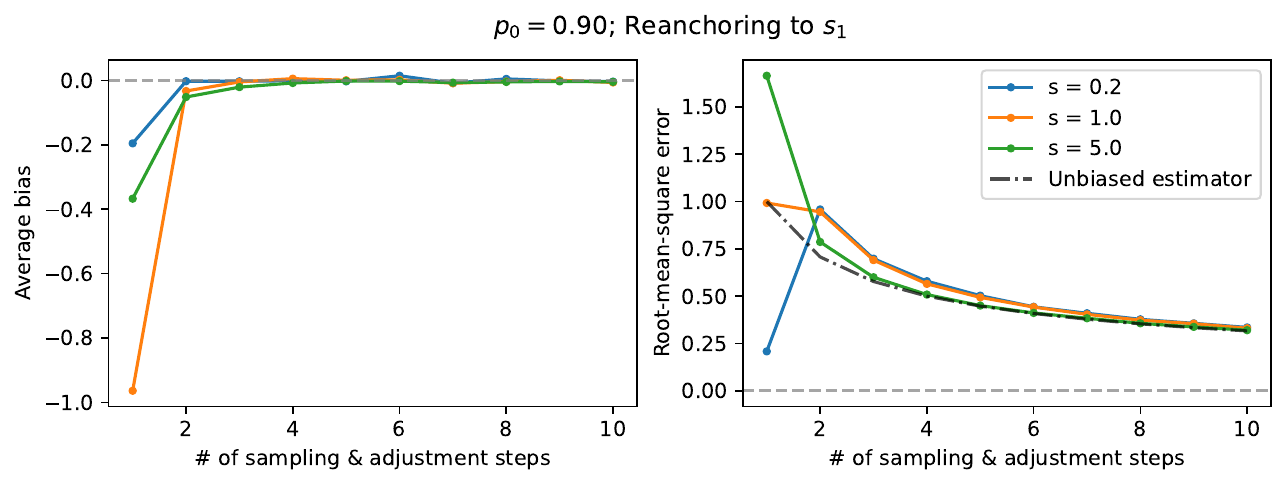}
\caption{ 
Bias and root-mean-square error as functions of the number of adjustment steps. If reanchoring sets $s_1$ (and $s_0$ in the spike) to the posterior mean $\hat{s}$, the anchor-induced bias persist for many steps (top panels). In contrast, if $s_1$ is used for reanchoring, the bias is reduced rapidly (bottom panels). Completely rational behavior corresponds to zero bias as the two options to chose from have the same expected value. 
The right-hand panels show that the 
root-mean-square error decreases as a function of the number of adjustment steps. This provides a principled explanation to the experimental finding, reported by~\cite{kirchler2017effect}, that fluctuations  shrins as a function of the number of steps.
Simulation parameters: bias $\sigma=1$, $p_0=0.9$, $s_0=0$, \# of samples: $10^4$.}
\label{fig:multistep_simulations}
\end{figure}

However, whatever fluctuations there are will be dependent on the number of steps you go through before issuing a verdict. The more time you have, the more refined an estimation you will do. In the Bayesian Anchoring framework, the number of adjustments you take in your process will affect how much variance there is in the verdict. Indeed, our simulations of multistep adjustment process (Fig.~\ref{fig:multistep_simulations}) confirm this intuition and show that the variance of the estimate decreases with the number of adjustment steps. This result shows that Bayesian Anchoring theory can qualitatively 
reproduce human risk behavior under time pressure and suggests that 
experiments similar to the ones performed in \cite{kirchler2017effect} 
could be designed to further validate the theory and pinpoint its exact details including the form of the reanchoring.

\begin{figure}
\centering
\includegraphics[width=1\linewidth]{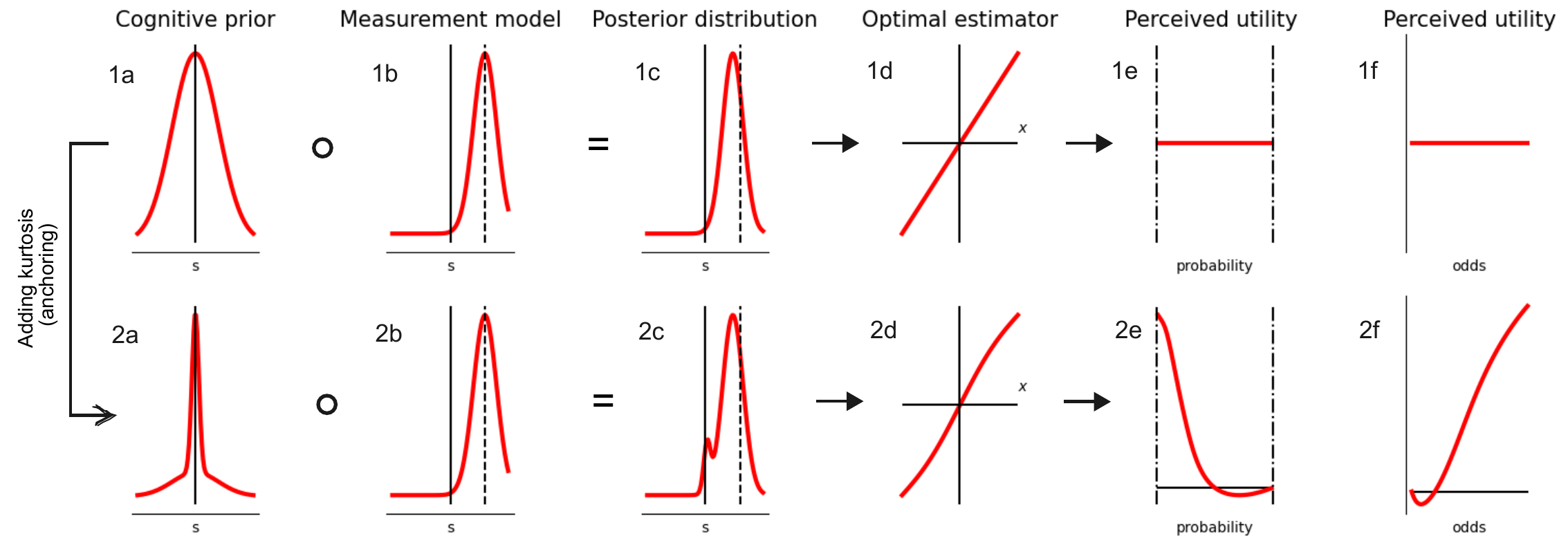}
\caption{
Key variables emerging from the theory, plotted in upper row  for a purely  Gaussian prior distribution and in the  lower row  for a leptokurtic prior. The different columns show (a) the prior distribution itself, mesokurtic in the upper row and leptokurtic in the lower row  2; (b) the internal measurement model, i.e. the relevant likelihood function, normal in both cases; (c) the resulting posterior distribution, which becomes bimodal in the leptokurtic case; (d) the optimal estimator used to evaluate utility, i.e. the mean of the posterior distribution; (e) the perceived utility as a function of the probability of a non-null outcome in a decision-making test such as those of~\cite{kirchler2017effect}; (6) the perceived utility, as a function of the odds  $q = \frac{1-p}{p}$. 
All results come from the gaussian mixture model defined  in Sec.  \ref{gaussian_mixture}), the parameter values  being   $\alpha =0,\gamma=2\sigma=1$ in the upper row and 
$\alpha =0.5,\gamma=2,\delta=0.3,\sigma=1$ in the lower row. 
 We fixed $x=3$ in panels b and c, $r =1$ in panels e and f of either row. }
\label{fig:gauss_vs_lepto}
\end{figure}

\section{Robustness and generalizations} 
\label{sec:general}
  
Most of the assumptions that went into the building of the theory are arguably natural -- such as the normality assumption on the model of inner measurement error (which leads  to  the  Gaussian likelihood in Eq.~\ref{Bayes_rule}). However, to model the prior distribution of the outcome 
we condensed experimental knowledge from the anchoring-and-adjustment literature into a specific functional form 
(Eq. \ref{priorslab}) that has been  rarely encountered in psychology so far, albeit being common in studies on numerical methods~\cite{mitchell1988bayesian,ishwaran2011consistency}. To test the robustness of the theory, we must check whether it would still be able to predict experimental observations if the prior distribution had a different form. 

Therefore, while so far we have deduced behavior from assumptions on the cognitive prior, we are now interested in the opposite logical path of inferring the underlying cognitive prior from what the experimental literature tells us on behavior (Fig. \ref{fig:flow}).  The detailed investigation we carry out in the Technical Supplement proves that indeed  a more general class of priors can account for the same fourfold pattern of risk attitudes, providing a  more nuanced model of resource-rational economic decision-making. However, all cognitive priors compatible with the fourfold pattern of risk attitudes  shares some essential features of the spike-and-slab prior. We report three rigorous results that clarify the connection between prior beliefs and behavior:
 
1. \textbf{Normality-rationality theorem}. 
All options in the experiment described by  Table \ref{table} correspond to the same expected gain/loss. In the language of  bounded rationality,  rational behavior would be to rank all options equally. A striking result of our framework is that normality equals rationality. As we prove in  Sec.~\ref{sec:gaussian_prior_yields_rationality}, rationality is equivalent to normality of the cognitive prior. A  subject's preference over options is rational only if the cognitive prior is a Gaussian distribution  (Fig. \ref{fig:gauss_vs_lepto}, panels 1e,1f). 

\begin{figure}
\centering
\includegraphics[width=.5\linewidth]{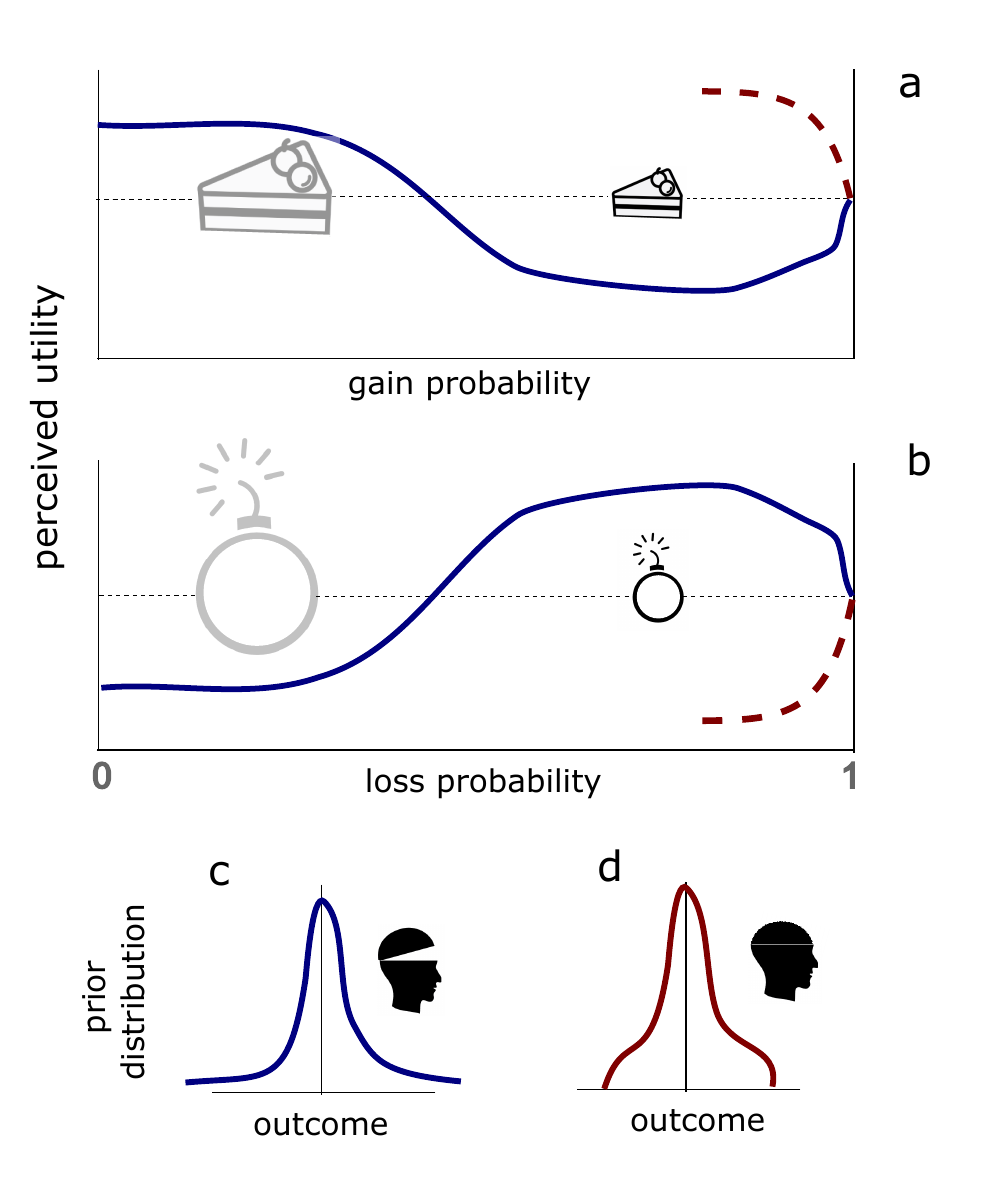}
\caption{
Sketch depicting the open mind theorem. 
If the prior has compact support as in panel d, the comparative utility function $U$ behaves as in the dashed red curves of panels a and b. An "open-minded" prior excluding no outcome a prior (panel c) is needed to give the perceived utility a slope compatible with the fourfold pattern (blue curve in panels a and b). 
}
\label{fig:open-mind}
\end{figure}

2. \textbf{Open mind theorem}. We show in Sec.~\ref{sec:open-mind} that  a compact-support prior is incompatible with experimental evidence. Since a Bayesian update cannot restore probability where it is zero at the outset, we would interpret a compact support as signifying that some potential values of the outcome have been ruled out a priori. Similarly, a non-compact support encodes an open-minded prior belief about possible outcome. Calculating the perceived utilities curves for our theory shows that, if the prior has compact support as in panel D
of Fig.~\ref{fig:open-mind}, the comparative utility function $U$ behaves as in the dashed red curves of panels A and B of Fig.~\ref{fig:open-mind}. The experimentally observed pattern of risk attitudes implies, therefore, that the prior beliefs are "open-minded". 

2. \textbf{Leptokurticity of the  cognitive prior}. As an additional result, the  emergence of the fourfold pattern reveals that the cognitive prior is  leptokurtic. Leptokurtic distributions can be hand-wavingly described as possessing thin tall peaks, and thick tails clearly separated from them~\cite{Pearson1905DASFU}. Thus, a  spike-and-slab prior is a caricature of what a leptokurtic distribution looks like in the unimodal case; such distributions are less restrictive than the spike-and-slab prior of Eq.~\ref{priorslab} but share  its essential features  (see e.g. Fig. \ref{fig:gauss_vs_lepto}, panel 2a). 
We can thus simply regard a leptokurtic prior as a "softer" way to encode anchoring. Any other type of prior distribution would not be able to yield the fourfold pattern of risk attitudes, as proven in Sec.~\ref{sec:prior_kurtosis} of the Technical Supplement.

In conclusion, the prior we used to encode the anchoring effect is not a necessary choice, but any cognitive prior compatible with experiments shares its essential features, which points back to anchoring-and-adjustment as a mechanism supporting or enhancing the fourfold pattern in fast decision making.   
 
%% ===========================================================
%% DISCUSSION
%% ===========================================================
\section{Discussion}

We introduced Bayesian Anchoring, 
a formalization of the anchoring-and-adjustment 
heuristic according to the Bayesian inference paradigm. 
In Bayesian Anchoring, the anchoring effect is implemented 
at the level of a prior belief utilized in the subsequent 
Bayesian inference that  is assumed to take place during  adjustment steps. 

In order to account for the experimentally observed features of 
the anchoring-and-adjustment process, 
the prior should strongly bias the inference towards the anchor. 
Arguably, the simplest prior satisfying this condition takes 
the form of a "spike and slab", which we adopted in our base analysis. 
We showed that this rather simple model reproduces 
the fourfold pattern of risk attitudes mechanistically and without the  somewhat arbitrary 
probability weighting introduced in Prospect Theory. We also showed that less stringent constraints on the initial prior lead to equivalent results.  The latest generation of decision-making experiments has provided new evidence not explained by Prospect Theory, such as  obtained by modulating processing speed and processing order. We showed that Bayesian Anchoring provides a mechanistic explanation for those results.

One of the central tenets of Prospect Theory is the idea that 
humans evaluate their options in terms of gains and losses with 
respect to a reference point, which depends on the context 
and can be manipulated from outside, for instance by 
rephrasing the description of a decision problem. 
Since losses and gains are not treated equally, 
the results of the evaluation depend on the reference point, 
leading to framing effects. 
Our results suggest that the existence of the reference point 
in Prospect Theory should be interpreted as an example 
of anchoring effect. 

The anchoring effect and other cognitive biases have been 
traditionally explained in terms of 
\textit{bounded rationality}~\cite{simon1955behavioral,kahneman2003maps} 
which posits that humans, given their imperfect knowledge 
of the world and limited cognitive resources, 
attempt to make good-enough ("satisficing") rather than optimal decisions. 
More recently this view has been enriched by quantitative models of cognitive 
biases in the framework of \textit{resource-rationality}~\cite{bhui2021resource}, 
wherein each unit computation has an associated cost and the optimal behavior 
has to balance costs of computation and sub-optimal decisions. 
In this framework, anchoring-and-adjustment has been interpreted as a 
costly MCMC sampling process \cite{lieder2012burn,lieder2018anchoring}. MCMC samplers start from an arbitrary initial guess and follow stochastic dynamics to generate samples from a desired distribution. 
Since the values generated in the initial phase are usually correlated 
with the initial guess and may not be representative of the desired 
distribution, practical implementations often incorporate a so-called 
\textit{burn-in} period during which values generated by the chain are 
discarded from the sampler. 
It has been shown~\cite{lieder2012burn,lieder2018anchoring} 
that "premature sampling", 
i.e. utilizing samples during the initial phase of the MCMC sampler, 
leads to biases that are similar to those generated due to 
anchoring-and-adjustment. Moreover, within this framework 
such a strategy of impatient sampling is often actually the 
optimal strategy.
In this context, let us reflect upon the role 
of a spike-and-slab prior in our theory as compared to 
its traditional applications in statistics, 
where it is usually introduced as an element of  
sparsity-inducing variable-selection techniques.  
We hypothesize that its normative role as used in the model 
of anchoring-adjustment is to provide a simple greedy technique 
of reducing the complexity of inference, as required by 
the rational utilization of limited computational resources. 
This view should be explicated by studying multi-step versions 
of our model in the context of resource-rationality, 
which we believe is an exciting avenue for future research.

Our model was inspired by the anticipated surprise introduced 
by Chan and Toyoizumi \cite{chan2021decision}. 
Their phenomenological model of economic decision making is based on 
the assumption that a decision maker maximizes the anticipated surprise 
defined as an expectation over a nonlinear function of the difference between 
the value of each outcome and the expected value across all outcomes. 
The form of the nonlinear function was chosen to qualitatively 
match experimental results. 
Our proposal can be interpreted as the normative 
underpinning of the anticipated surprise model, 
although this mapping comes with a caveat: 
the original anticipated surprise model assumes 
a convex nonlinear function for positive arguments, 
which the optimal estimator $\hat{s}$ is not 
(see Sec.~\ref{sec:spike_slab} of the technical supplement, 
or Fig.~\ref{fig:gauss_vs_lepto}, panel B4). 

In deriving the results of section 
\ref{sec:general}, the finding about the posterior (such as leptokurticity) were translated into constraints on the prior by assuming a perfectly Bayesian framework. Such deductions would have been completely different if we had assumed a mechanism to go from prior to posterior that includes costs for the process itself of hypothesis update, as in Ref. \cite{prat2021biases}. Such theories of "Costly Bayesian inference" may have the potential to yield the observed behavior with different priors. Identifying the most plausible way to explain observations with theories of this kind remains a problem open to exploration.  

Another future direction concerns the exploration of different anchoring mechanisms. The model assumed that the initial guess (the anchor) coincides with the risk-free option if available. However, there are settings  where such an anchor may not not available, and the gathering of experimental data on such settings is a necessary next step to calculate the prior distributions that lead to best reproducing empirical results. 

Finally,  the vanilla version of Bayesian Anchoring presented here does not account for the overall risk aversion for losses observed in human decision making. This will also constitute a  motivation for additional research on Bayesian Anchoring. 

\section{Acknowledgments}

We would like to thank  Taro Toyoizumi for introducing us to the problem, Ho Ka Chan  for discussions on prospect theory and its alternatives, and Manuel Baltieri for inspiring conversations on the range of applicability of Bayesian frameworks. 

\bibliography{library}
%% ========================================================
%% SUPPLEMENTARY INFORMATION
%% ========================================================
\newpage
  
\onecolumngrid% or \widetext

\setcounter{page}{0}
\pagenumbering{arabic}
\setcounter{page}{1}
\clearpage
 
\begin{center}
\textbf{TECHNICAL SUPPLEMENT}
\end{center}

\renewcommand{\thefigure}{S\arabic{figure}}
\setcounter{figure}{0}

\setcounter{equation}{0}
\setcounter{figure}{0}
\setcounter{table}{0}
\setcounter{page}{1}
\setcounter{section}{0}
\makeatletter
\renewcommand{\thesection}{S-\Roman{section}}
\renewcommand{\theequation}{S\arabic{equation}}
\renewcommand{\thefigure}{S\arabic{figure}}

\section{Notation} 
\label{sec:notation}

This supplement contains mathematical derivation of various aspects of the theory helpful for a fuller  comprehension of its structure. To ease the writing and reading of such derivations, in the  supplement we will use a leaner notation than in the main text. 

\begin{itemize} 

\item  The function $g_\sigma(x)$  is, as in the main text, a standard Gaussian probability density with zero mean and standard deviation $\sigma$. To  avoid  lengthy pedices, where  the expressions for  variances get long we will use the notation $G(x;\sigma) =  g_{\sigma}(x)$  and $G_{\mu \sigma}(s) = g_\sigma (s-\mu)$.

\item The prior distribution will be called simply $\rho(s)$ rather than  $\rho_0(s)$ as in the main text. 

\item Random utilities are measured as shifts from the mean value of their prior distribution (equal to the anchor $s_0$ for the spike-and-slab case), so that the prior itself has zero mean. The variable $s$ (now meaning $s- s_0$) will be referred to as the "deviation", and "zero-deviation" will mean "for $s=0$. We will call $x$ the displacement of the outcome from that mean, which in the main text would have been called $x- s_0$.  

\item The posterior distribution conditioned to a given value $x$ of the measurement will be called $\phi(s|x)$ (with its dependence on the measurement error $ \sigma$ being implicit) rather than $\rho(s |(x,\sigma))$. 

\item We use overlines to indicate averages over $\rho$: 

\be \overline{ h(s) }  = 
\int_{-\infty}^{\infty} ds \ h(s) \rho(s) \ee

\item We  use the notation $  \langle \ldots \rangle_{x}$ (or simply $\langle \ldots \rangle$, leaving the value of the outcome implicit) to denote the average over the posterior given outcome $x$. 

\item Finally, assuming the measurement error to be  Gaussian with variance  $\sigma^2$, we write the likelihood $\rho((x,\sigma)|s) $ as $ g_\sigma(x-s )$. 

  \end{itemize}
The one-step Bayesian update is thus described by: 

\be
\label{posterioransatz}
\phi(s|x)  = 
\frac{\rho(s) g_\sigma(x-s ) }{
\int ds \ \rho(s) g_\sigma(x-s )}
\ee

\section{General properties} 

\subsection{Symmetry of the prior }
\label{sec:symm_of_prior}

With the extra assumption that $\rho(s)$ is symmetric around 0, it follows  that the optimal estimator given by the posterior mean $\hat{s}(x) = \langle s \rangle_x $  is antisymmetric.

Indeed write 

\be \hat{s}(x)  =  \frac{ \int ds  \ s \rho(s) g_\sigma(x-s) }{ \int ds \rho(s) g_\sigma(x-s) } \ee

and notice that if we replace x with -x, but then flip the sign of the integration variable and use the symmetry of $g_\sigma$ , we end up having on the RHS the same object as before but with a minus in front.
   
The converse is also true. Indeed using the identity 

\begin{eqnarray}
\label{gauss_derivatives_identity}    
 \partial_x g_\sigma(x-s) 
&=& 
\sigma^{-2} (s-x)g_\sigma(x-s)
\end{eqnarray}

we can write 

\be
\label{f_decomposition} 
\hat{s}(x) = x + \sigma^2 \partial_x \log \int ds \ \rho(s) g_\sigma(x-s)  = 
x + \sigma^2 \partial_x \log \left[\exp\left( \frac{\sigma^2 \partial^2_x}{2} \right) \rho(x)\right] 
\ee

where the second term represents the mean-value shift from measurement due to the prior's traction, expressed compactly via the differential formulation of a Weierstrass transform~\cite{hirschman2012convolution}. 
 
As seen from~(\ref{f_decomposition}),  antisymmetry of $\hat{s}(x)$ implies antisymmetry of the second term on the RHS, hence symmetry of the function subjected to  $\partial_x$. Log being bijective on $\mathbb{R}^+$, that entails in turn symmetry of the argument of the logarithm, that is, of the Weierstrass transform of the prior. But the Weierstrass transforms commute with the parity operator, from which we infer  symmetry of the prior itself.

\subsection{Gaussian identities} 
\label{sec:gaussian_case}
 
For some of the following manipulations it will be useful to  recall, as a preliminary,  that the product of two Gaussian densities can be written as 

\begin{equation}
\label{gaussianproduct}
G_{\mu_1, \sigma_1}(s)  G_{\mu_2, \sigma_2}(s) =
G_{0, \sqrt{\sigma_1^2+\sigma_2^2}} (\mu_1-\mu_2)  
\ G_{\mu_{12},\sigma_{12}}(s) 
\end{equation}

where

\be \mu_{12} = \frac{\mu_1 \sigma_2^2 + \mu_2 \sigma_1^2 }{\sigma_1^2 + \sigma_2^2}
\hspace{30pt}
\sigma_{12}^2  = \frac{\sigma_1^2 \sigma_2^2}{\sigma_1^2 + \sigma_2^2} \ee 

So if the prior is Gaussian with zero-mean and variance $\gamma^2$,  applying~(\ref{gaussianproduct}) with $\sigma_1=\sigma$ and $\sigma_2=\gamma$ we see that the resulting normalization prefactor is $g_\Sigma(0)$, where 
we are establishing the notation 
$ \Sigma^2 = \gamma^2 + \sigma^2$, 
 and the normalized posterior distribution is $\phi(s | x ) = G_{\hat{s}(x),\Gamma}(s)$
with the posterior variance 
 
 \be
 \label{gaussian_posterior_var}
 \Gamma^2 = \frac{\sigma^2 \gamma^2}{\sigma^2 + \gamma^2}
 \ee
 
 (note that $\sigma \gamma = \Sigma \Gamma$) and  the posterior  mean
  
\be
\label{gaussian_mean}
\hat{s}(x) \equiv \langle s |x  \rangle = \langle s  \rangle_x = 
\frac{\gamma^2 x }{\sigma^2 + \gamma^2}
\ee
 
which is linear in $x$. In particular, it will be useful to remember that the posterior mean is  $\mathcal{O}(x)$ for large $x$ whereas the variance is $\mathcal{O}(1)$.
 
\subsection{Connection between perceived utility and  posterior cumulants} \label{sec:cumulants_theorem}

We are going to prove the statement that  

\be
\label{cumulant_relation}
\partial^n_x \hat{s}(x) = \sigma^{-2n} \mathcal{C}_x^{(n+1)}
\ee

where we defined the cumulants of the posterior distribution as  

\be
\label{cum_def} \mathcal{C}_x^{(n)} = n! \left[ \frac{\partial^n}{\partial t^n} \log \langle 
 e^{ts}\rangle_{s|x} \right]_{t=0}
\ee 

We will prove relationship~(\ref{cumulant_relation}) by induction. For $n=0$, (\ref{cumulant_relation}) reduces to $\hat{s}(x) = \langle s \rangle_{s|x} $ and therefore is true. 
Assume  that Eq. (\ref{cumulant_relation}) is verified  for  $n-1$. We write then its LHS for $n$ as

\be
\label{induction}
\partial^{n}_x \hat{s}(x) = 
\partial_x
\left( \partial^{n-1}_x \hat{s}(x) \right) = 
\partial_x
\left( \sigma^{-2n+2}
\mathcal{C}_x^{(n)}\right) 
=
 \sigma^{-2n+2}
\partial_x
\mathcal{C}_x^{(n)}
\ee 

To calculate this derivative of a cumulant, we begin by applying  identity~(\ref{gauss_derivatives_identity})
to the derivative of the cumulant generator

\be
\label{der_log}
\partial_x \log \langle e^{ts} \rangle = \sigma^{-2} \frac{ \langle (s-x) e^{ts} \rangle_{s|x} }{\langle e^{ts}\rangle_{s|x}} = -\sigma^{-2} x + \sigma^{-2} \partial_t \log \langle e^{ts} \rangle =\sigma^{-2} \partial_t \sum_{n\geq 0}
\frac{\mathcal{C}^{(n)}}{n!} t^n +h(x)
=\sigma^{-2}\sum_{n\geq 0}
\frac{\mathcal{C}^{(n+1)}}{n!} t^n +h(x)
\ee
where $h(x)$ is not a function of $t$ and therefore has no effect on derivatives with respect to $t$. 

The derivative of the cumulants is calculated by  applying Eq.
(\ref{cum_def}) and then Eq. ~(\ref{der_log}). We find that, for all $n \geq 1$,

 \be
\label{der_cum}
\partial_x \mathcal{C}_x^{(n)}
= n! \left[ 
\partial_t^n \partial_x 
\log \langle e^{ts} \rangle \right]_{t=0}
=
\sigma^{-2} \mathcal{C}^{(n+1)}
\ee

while explicit calculation shows that
equation (\ref{der_cum})
is also valid for $n=0$. 

Substituting relationship (\ref{der_cum}) into Eq. (\ref{induction}) leads straight to formula (\ref{cumulant_relation}),  thus proven by induction. 
 
\subsection{Range of the posterior mean} 
\label{sec:post_mean_range}
 
We show  here that $\langle s \rangle_x$ lies between $\bar{s}$ and $x$ if the prior distribution is symmetric and unimodal. 

Under these conditions, we can write the $\rho(s)$ in terms of a zero-centered distribution nonincreasing for positive argument, as $ \rho(s) = \hat{\rho}(s - \bar{s})$.  We first notice that, defining the normalization factor $\mathcal{N}  =  
\int ds  g_{\sigma}
(x-s) \rho(s) $, we have

\be 
\hat{s}(x) - x  = \mathcal{N}^{-1} \int ds  (s-x) g_{\sigma}
(x-s) \rho(s) = 
\mathcal{N}^{-1}   \int dy \ y \ g_\sigma(y) \ \hat{\rho}(y + \hat{x})
 = 
 \mathcal{N}^{-1}  \int_0^\infty dy \ y \ g_\sigma(y) \ \Big( \ \hat{\rho}(\hat{x} + y)  -
\hat{ \rho}(\hat{x}-y) \Big) 
\label{zero_integral} 
\ee

where $\hat{x} = x - \bar{s}$. It's clear by symmetry that this integral vanishes for $\hat{x}=0$. Without lack of generality, let's check its sign for $\hat{x} > 0$. We have then

\bea
\hat{s}(x) - x  &=& \mathcal{N}^{-1}  \Bigg[ \int_0^{\hat{x}}   \Big( \hat{\rho}(\hat{x}+y) - \hat{\rho}(\hat{x}-y) \Big) 
+ 
\int_{\hat{x}}^\infty  \Big( \hat{\rho}(\hat{x}+y) - \rho(\hat{x}-y) \Big) 
\Bigg] 
y g_\sigma(y) dy 
\\ 
&=&
\mathcal{N}^{-1}   \int_0^{\hat{x}}   \Big[ \hat{\rho}(\hat{x}+y) - \hat{\rho}(\hat{x}-y) \Big] 
y g_\sigma(y) dy 
+ 
\mathcal{N}^{-1}  \int_0^\infty  \Big[ \hat{\rho}(2\hat{x} + v) - \hat{\rho}(-v) \Big]  (v+\hat{x}) g_\sigma(v+\hat{x}) dv
\label{fx_minus_x}
\eea

The sign of this expression follows from the assumption that  $\rho(s)$ is symmetric and  monotonous on either side of its mean. Indeed, in the first term we notice $\hat{x}+y>\hat{x}-y>0$ and thus
$\hat{\rho}(\hat{x}-y) > \hat{\rho}(\hat{x}+y)$ and in the second term we notice $ \rho(-v) = \rho(v) > \rho(2 \bar{x} + v) $. Thus, in both terms, the  square brackets have nonpositive  content and multiply positive factors, leading  to the conclusion that expression~(\ref{fx_minus_x}) is nonpositive, hence 
$\hat{s}(x) \leq x$. Analogously, $\bar{x} < 0$ leads to the result $\hat{s}(x) \geq x$, i.e. $\hat{s}(x)$ lies on the same side of $x$ as $\bar{s}$ does. 

To prove that, conversely,  $\hat{s}(x)$ lies on the same side of $\bar{s}$ as $x$, one proceeds as above with the roles of likelihood and prior swapped, exploiting the monotonicity of the Gaussian density on either side of its mean.

\subsection{Bounds on the derivative of the posterior mean} 
\label{sec:post_mean_derivative}

The posterior mean can be written as $ \hat{s} = \overline{s \mathcal{L}} / \overline{\mathcal{L}}$ in terms of prior averages, where $\mathcal{L}$ is an arbitrary likelihood $\mathcal{L}(s) := \rho((x,\sigma)|s)$. Its derivative is therefore

\be
\label{ratio_general}
\hat{s}' = \frac{ \overline{\mathcal{L}} \ \overline{s \mathcal{L}'} -\overline{s \mathcal{\mathcal{L}}} \ \overline{\mathcal{L}'} }{\overline{\mathcal{\mathcal{L}}}^2}  
\ee

The  assumption  of a Gaussian measurement model turns Eq.~(\ref{ratio_general}) into the ratio between the variance of the posterior and the variance of the likelihood

\be
\label{var_ratio}
\hat{s}'(x) = \frac{\textrm{variance of posterior for  outcome $x$}}{ \textrm{variance of likelihood}}
\ee

as also follows from  Eq.~(\ref{cumulant_relation}) evaluated with $n=1$. 

Because this derivative is in principle measurable from experiments that probe the perceived utility curve, its range of  values may offer a further experimental test for the theory. Obviously it can never be equal to zero, because the numerator of Eq.~(\ref{var_ratio}) vanishes only if the observations are completely uninformative or the prior itself has zero variance. This marks a difference between the theory presented here  and the power-law ansatz seen in \cite{chan2021decision}, where this derivative vanishes in the origin. 

Under the assumptions of 
Sec.~\ref{sec:post_mean_range}, it can be seen that the ratio (\ref{var_ratio})  is always subunity at $x=0$. To prove that, it is sufficient to restate the main result of Sec.~\ref{sec:post_mean_range} as $\frac{\hat{s}(x)}{x} <1$. Since under  the same conditions  $\hat{s}(0)=0$, we find 
 
\be 
\lim_{x \rightarrow 0} \hat{s}'(x) 
= 
\lim_{x\rightarrow 0} \frac{\hat{s}(x)}{x} < 1 
\ee

\subsection{Asymptotics of the posterior mean}
\label{sec:post_mean_asympt}

The asymptotics of the posterior mean has asymptotic scaling contained between $O(1)$ and $O(x)$ in the limit of large outcomes ($x\rightarrow \infty$). 

To see this, begin by noticing that the posterior mean is a nondecreasing function of the outcome $x$. From this, it  follows  that  no scaling below $O(1)$ is possible.

On the other hand, suppose the scaling was larger than $O(x)$. It would then be possible to find a sufficiently large value  $x*$, such that for $x>x*$ the inequality $\langle s \rangle \geq x $ is violated. Since  this inequality was proven in Sec.~\ref{sec:post_mean_range}, it follows that ( under the same conditions given in that section) the large-$x$ scaling of $\hat{s}(x)$ is linear at most. 

\subsection{Compact-support priors and  posterior mean}
\label{sec:compact-support-and-posterior-mean}

A preliminary step to prove the open-mind theorem 
is to study the asymptotic behavior of the posterior mean for compact-support priors. 

In the following we consider the limit $x \rightarrow \infty$ for convenience, although the same considerations apply to the limit $x \rightarrow -\infty$.  We prove that the  two statements of  $\hat{s}(x)$ having  a finite limit for large $x$ and the  prior having  compact support are equivalent, as long  as 
$ \rho \in C^0$ and   $ \sigma < \infty$. 

We start from the assumption that the  prior has compact support. The set of points $\{ s : \rho(s)>0\}$ has therefore a  supremum $S < \infty$, and from the definition of $\hat{s}(x)$ it follows $\hat{s}(x) < S \ \forall x$, entailing that the large-x limit of $\hat{s}(x)$ is finite. 

Vice versa, to prove that a finite limit entails a compact support, we suppose ad absurdum that the support is not compact. It follows that for every $L\in\mathbb{R}$  we can find  an increasing sequence $s_1<  s_2 < \ldots$ such that $L<s_1$ and  $\rho(s_i)>0 \ \forall i \geq 1$. Since $\rho \in C^0$, we also have  that $\rho(s)$ is positive in a neighborhood of each $s_i$. 

The statement being true for any $L$, it   holds in particular for $L= \lim_{x \rightarrow \infty} \hat{s}(x) $. We  write therefore 

\be 
\label{weighted_av}
\hat{s}(x) =  \langle s  \rangle_{s|x}^{[-\infty,s_1]} w(-\infty,s_1 |x) 
+  \langle s  \rangle_{s|x}^{[s_1,\infty]} w(s_1,\infty|x) 
\ee
where 
\be
\label{average_over_interval}
\langle s \rangle_{s|x}^{[a,b]}  = 
\frac{ \int_a^b ds \ s  \ \rho(s) g_\sigma(s-x) }{\int_a^b ds   \ \rho(s) g_\sigma(s-x)} 
\ee
\be 
w(a,b|x) =  \frac{ \int_a^b ds \  \rho(s) g_\sigma(s-x) }{\int_{-\infty}^\infty ds   \ \rho(s) g_\sigma(s-x)}  \ee 
where $\rho(s)$ being positive in a right neighborhood of $s=s_1$ results in $w(M,\infty|x)>0$ . 

Given  that formula~(\ref{average_over_interval}) is an average over the interval $[a,b]$ with the distribution 

\be \psi(s; a,b) = \frac{\rho(s) g_\sigma(s-x) \theta(s-a) \theta(b-s)}{\int_a^b ds   \ \rho(s) g_\sigma(s-x)}  \ee

it follows that 

\be
\label{ineq_averages}
\langle s \rangle_{s|x}^{[s_1,\infty]} 
\geq \langle s \rangle_{s|x}^{[-\infty,s_1]}  
\ee

Let us now find bounds for the weights $w$'s. We note  that 

\be 
\label{w_upper_bound}
w(-\infty,s_1|x) 
\leq
\Big( \int_{-\infty}^M 
\rho(s) ds \Big) \ \mathrm{sup}_{s\in (-\infty, s_1)} g_\sigma (s-x) = 
g_\sigma(s_1-x) \int_{-\infty}^M 
\rho(s) ds
\ee 

where the last equality holds for  sufficiently large $x$  ($x \geq s_1$). We also note that there is a right neighborhood of $s_2$ with  size $\delta$ where $ \rho(s) >0$, hence 

\be 
\label{w_lower_bound}
w(s_1, \infty|x)
\geq 
 \frac{ \int_{s_2}^{\infty} ds \  \rho(s) g_\sigma(s-x) }{\int_{-\infty}^\infty ds   \ \rho(s) g_\sigma(s-x)}  
 \geq 
\frac{ g_\sigma(s_2-x) \int_{s_2}^\infty \  ds \  \rho(s)  }{\int_{-\infty}^\infty ds   \ \rho(s) g_\sigma(s-x)}  
 \ee
   
where the second inequality holds for sufficiently large $x$ ($x \geq s_2$).

From Eqs.~(\ref{w_upper_bound}) and 
(\ref{w_lower_bound}), it follows that 
  
 \be 
 \label{w_ratio}
\lim_{x \rightarrow \infty} \frac{ w(-\infty,s_1|x)}{w(s_1, \infty|x)} 
\leq
\frac{ \int_{-\infty}^{s_1}
\rho(s) ds}{
\int_{s_2}^\infty 
\rho(s) ds}
\lim_{x \rightarrow \infty} 
\frac{g_\sigma(s_1-x)}{g_\sigma(s_2-x)} = 0 
\ee
 
On the other hand, for any $x$ we have 
 $  w(-\infty,s_1|x) + 
w(s_1,\infty|x)
=1 $, 
which together with Eq. (\ref{w_ratio}) yields 

\be
\label{w_limits}
\lim_{x\rightarrow \infty} w(-\infty,s_1|x) = 0 \hspace{40pt} 
  \lim_{x\rightarrow \infty} w(s_1,\infty|x) = 1 
 \ee
 
Applying ~(\ref{w_ratio}) and  
Eq.~(\ref{ineq_averages}) 
 into Eq.~(\ref{weighted_av}), we obtain

\be \lim_{x\rightarrow \infty} \hat{s}(x) = 
\lim_{x\rightarrow \infty}    \langle s \rangle_{s|x}^{[s_1,\infty]}   w(s_1,\infty|x) = \lim_{x\rightarrow \infty}   \langle s \rangle_{s|x}^{[s_1,\infty]}    
\ee

On the other hand, from Eq.~(\ref{average_over_interval}) we have $ \langle s \rangle_{s|x}^{[s_1,\infty]} \geq s_1 > L$, so we reach the contradictory conclusion that $ \lim_{x\rightarrow \infty} \hat{s}(x) > L$, which shows that a finite limit  $L$ cannot exist.

\section{Definition of the perceived utility function}
\label{sec:perceived_utility}

The experimental setting used to verify the Reflection Effect~\cite{kahneman1979prospect} is described by Table~\ref{table} of  the main text. The function that determines the \textit{perceived} utility given a certain  outcome is  $  \hat{s}(x)=\langle s \rangle_x$. This function  must be averaged over all outcomes to yield the actual value of the perceived utility $U$ associated to any choice. 

Therefore, the perceived utility  becomes  

\be 
U  (p,r) = U_A(p,r) - U_B(p,r) = p \hat{s}(r/p -r) + (1-p) \hat{s}(-r)
\ee

and is written most concisely in terms of the odds $q = \frac{1-p}{p}$ (ranging between $0$ and $\infty$),

\be 
\label{utility}
U(q,r) = \frac{ \langle s \rangle_{q r} + q \  \langle s \rangle_{-r}}{1 + q} 
\ee

The case of a risk-free outcome is recovered for $q=0$, when indeed we have $U_A =U_B $ and thus $U(q,r) = 0$. 

In the main text we started out by deriving a theory based on a specific cognitive prior (the spike-and-slab prior), and calculated from it a specific form of the function~\ref{utility}, showing that it leads to  experimentally verifiable results. In this supplement, we would like to proceed the other way around. Namely, we will now make some experimentally motivated demands on the perceived utility function given in Eq.~\ref{utility} (which constitutes of course the final output of our theory) and will asking what cognitive priors would output a perceived utility fulfilling those requirements. 

The reflection effect amounts to the requirement that perceived utility $U$ flips sign as a function of the gain/loss probability $p$, becoming positive for sufficiently large $p$ if $r$ is negative, and becoming  negative for sufficiently large $p$ if $r$ is positive. 

The goal therefore  is to characterize the set of prior densities $\rho(s)$ such that: 

(A) For any $r>0$, the equation $ U(q,r)= 0$ is satisfied for a single $ q \in (0,\infty)$, let's call it $q_+$. At  $q_+$ the lowest-order nonvanishing derivative of $ U$ with respect to $q$ is positive. 

(B) For any $r<0$, the equation 
$ U(q,r)= 0$ is satisfied for a single $ q \in (0,\infty)$, let's call it $q_-$ . In $q_-$, the lowest-order nonvanishing derivative of $ U$ with respect to $q$ is negative. 

The point that makes $U$ vanish will be called in either case the choice-switching point. Experiments suggest that $q_+ \approx q_-$ (reflection principle) and that this is around $\approx 0.2$. 

Assuming $\rho$ symmetric makes $\hat{s}(x)$ antisymmetric, and a consequence of that is that conditions (B) are automatically satisfied if conditions (A) are.   

\section{Particular cases}
 
\subsection{Spike-and-slab prior}
\label{sec:spike_slab}

Although in this supplement we do not restrict ourselves to the spike-and-slab form of the prior, that specific prior deserves special attention because it encodes relevant experimental knowledge from anchor-and-adjust heuristic. We review it briefly in this section. 

The spike-and-slab prior already stated in the main text (Eq.~\ref{priorslab}) is 

\be 
\rho(s) = p_0 \delta(s) + (1-p_0) g_{\gamma}(s) 
\ee

and leads to a posterior distribution of the form  

\be \phi(s) \propto
p_0 g_{\sigma}(x) \delta(s) + (1-p_0) g_{\sqrt{\gamma^2+\sigma^2}}(x) 
g_{\frac{ \gamma \sigma}{ \sqrt{ \gamma^2+\sigma^2}}} \left( s - \frac{\gamma^2}{\gamma^2+\sigma^2}  x \right) 
\ee

The resulting optimal estimator (posterior mean) is then

\begin{equation}
\label{s_hat_spike_slab}
\hat{s} (x)
= 
\frac{a x}{1 + b \exp\left(-\frac{x^2}{2 \tilde{\sigma}^2}\right)},
\end{equation}

with positive constants $a,b,\tilde{\sigma}$ given by 

\be a = \frac{\gamma^2}{\gamma^2 + \sigma^2}, \hspace{10pt}
 b = \frac{p_0}{1-p_0}\frac{\sqrt{\gamma^2+ \sigma^2}}{\sigma}, \hspace{10pt} \tilde{\sigma} = \frac{\sigma}{\sqrt{a}} \ee

For $x \ll \tilde{\sigma}$ this reduces to

\be
\hat{s} (x)
\approx
\frac{a}{1+b} x + \frac{a b}{(1+b)^2}\frac{x^3}{2\tilde{\sigma}^2}
\ee

which demonstrates that, around $x=0$, this estimator features a 
 shape that is convex for positive $x$'s and concave for negative $x$'s, suppressing small 
fluctuations. However the convex-concave shape is lost for larger values of $|x|$, and for large $x$ the estimator approaches a linear  asymptote $\hat{s} = a x$ from below.

Plugging into Eq.~(\ref{utility}), we obtain 

\be
U(q,r) = \frac{a q r }{1 + q } \left(
\frac{1}{1 + b \exp\left( - \frac{  q^2 r^2}{  2 \tilde{\sigma}^2 }\right) }
- 
\frac{1}{1 + b \exp\left( - \frac{   r^2}{  2 \tilde{\sigma}^2 }\right)}
\right)
\ee

which is the function shown in Fig.~\ref{fig:perceived_utility} of the main text.

\subsection{Gaussian-mixture prior} 
\label{gaussian_mixture}

By a gaussian-mixture prior we mean a linear combinations of two gaussian density functions.As seen in the main text, a  gaussian-mixture prior with one infinitely narrow gaussian ("spike") fullfills the phenomonological requirement for any value of $\sigma$.  On the other hand, assuming Gaussian priors are an adequate model, we don't expect that the value of $\sigma$ in the actual experiment would ever vanish, and we may assume that in all experiments conducted so far $\sigma$ was larger than some baseline value $\delta$. Such a gaussian-mixture prior has been used to generate figure \ref{fig:gauss_vs_lepto}, and for this reason we see it worthwhile to outline it here. 

We write 

\be
\label{gaussian_mix_prior}
\rho(s) = \alpha g_\delta(s) + (1-\alpha) g_\gamma(s) 
\ee

for a constant $\alpha \in ]0,1[$, with $ \gamma > \delta$.  

Since the prior \ref{gaussian_mix_prior} is symmetric, all the odd cumulants vanish. Let us compute the central moments 

\bea
\overline{s^2} &=& (1-\alpha)\gamma^2 + \alpha \delta^2  < \gamma^2
\\ 
\overline{s^4} &=& 3 (1-\alpha) \gamma^4  + 3 \alpha \delta^4 < 3 \gamma^4
\\
\eea 

and the kurtosis 

\bea 
\kappa &=&  \overline{s^4}  \big/ (  \overline{s^2} )^2= 
\frac{3}{\alpha}
\frac{1 + \frac{1-\alpha}{\alpha} \eta^4}{
1 + 2 \frac{1-\alpha}{\alpha} \eta^2 + \frac{(1-\alpha)^2}{\alpha^2} \eta^4}
\eea

where $ \eta\equiv\gamma/\delta$. Since  $\kappa > 3$ for any $ \eta \geq 0 $, this is a leptokurtic prior.  

Applying Eq.~\ref{gaussianproduct} leads to the posterior distribution

\be \phi(s) = (1-\beta(x)) G\left(s - \frac{\gamma^2 x}{\sigma^2+\gamma^2}; \frac{\gamma\sigma}{\sqrt{\sigma^2+\gamma^2}}\right) + 
\beta(x) \  G\left(s - \frac{\delta^2 x}{\sigma^2+\delta^2}; \frac{\delta\sigma}{\sqrt{\sigma^2+\delta^2}}\right) 
\ee 

where 

\be \beta(x) = 
\frac{ \alpha G( x; \sqrt{\sigma^2 + \delta^2})}{
(1-\alpha)  G\left(x;  \sqrt{\sigma^2 + \gamma^2}\right)+ \alpha G\left(x;  \sqrt{\sigma^2 + \delta^2}\right)}
\ee 

Applying Eq.~\ref{gaussianproduct} leads to the optimal estimator 
 
\be
\hat{s}(x)  = 
\frac{ \frac{(1-\alpha) \gamma^2}{(\sigma^2 + \gamma^2)^{3/2}} 
e^{  \frac{ x^2}{2 \lambda^2}  } 
+ 
 \frac{\alpha \delta^2 }{(\sigma^2 + \delta^2)^{3/2}} 
}
{
\frac{1-\alpha}{(\sigma^2 + \gamma^2)^{1/2}}  
e^{  \frac{ x^2}{2 \lambda^2}  } + 
\frac{\alpha}{(\sigma^2 + \delta^2)^{1/2}}
 }  \ x 
\ee

where 

\be 
\lambda^2 = \frac{(\delta^2 + \sigma^2)(\gamma^2+\sigma^2)}{\gamma^2 - \delta^2}.
\ee

From this, applying 
Eq.~\ref{utility},  we see that the formula for perceived  utility  has the form 

\begin{equation}
U  (q,r)  = K r 
\ 
\frac{
e^{-\frac{ q^2 r^2}{2 \lambda^2}} - e^{- \frac{ r^2}{2 \lambda^2} }}{
A
e^{-\frac{(q^2+1)r^2}{2 \lambda^2}}
+
B
e^{-\frac{q^2 r^2}{2 \lambda^2}}
+
B
e^{-\frac{r^2}{2 \lambda^2}}
+
C
} 
\label{Delta_with_delta}
\end{equation}

where 
  
\bea 
K &=& \alpha (1 + \alpha) \frac{ q}{1 + q}
\frac{\sigma^2 (\gamma^2 - \delta^2)}{\sqrt{\sigma^2 + \gamma^2}\sqrt{\sigma^2 + \delta^2} } \\
A &=& (1- \alpha)^2 (\sigma^2 + \delta^2)\\
B &=& \alpha (1 - \alpha) \sqrt{\sigma^2 + \delta^2}
\sqrt{\sigma^2 + \gamma^2} \\
C &=& \alpha^2( \sigma^2 + \gamma^2)
\eea
 
This indeed has a single zero at the half-point ($q=1$). As seen from the only part of the numerator that changes sign, at that point the functions turns from negative to positive if $r>0$, so  it is indeed an increasing function of q there  (and  a decreasing one if $r<0$)
as desired.

All of the above formula also hold true in the limit $ \delta \rightarrow 0$, where one of the two gaussians becomes a delta function and the spike-and-slab results are recovered with $p_0  = \alpha$.

\section{General results}

\subsection{Posterior mean vanishes for zero deviations}
\label{sec:mean_vanishes}

The reflection effect is, in our framework, the fact that the perceived utility  $U(p,r) $ gets positive for sufficiently large $p$ if $r$ is negative, and gets negative for sufficiently large $p$ if $r$ is positive. From that fact, we infer in particular the inequalities 

\be  U(p=1,r>0) \geq  0 \ee
\be  U(p=1, r<0) \leq 0 \ee

On the other hand, from the definition of $U(p,r)$ we see that $U(p=1,r)$ is always equal to $\langle s \rangle_0 $ regardless of the sign of $r$. So the only way to satisfy both inequalities is if $ \langle s \rangle_0 = 0 $. 

That's of course automatically satisfied if the prior is symmetric around zero. But as seen here, it has to be true also for asymmetric priors.

\subsection{Gaussian prior generates rational preferences}
\label{sec:gaussian_prior_yields_rationality} 

Eq.~(\ref{gaussian_mean}) shows that the posterior mean is linear in the measurement.

Plugging that linear law into Eq.~(\ref{utility}), we find  
\be U(q,r) \equiv 0 
\label{U}
\ee
for any value of $q$ and for any value of $r$, as shown in Fig.~\ref{fig:gauss_vs_lepto}. 
 
The finding is therefore that $U$ is a measure of nongaussianity -- or, in other words,the Kahneman-Tversky  experiment itself measures the nongaussianity of the prior. If the prior was  Gaussian, experimental subjects would choose indifferently among options with any risk level. Thus, the "linearity in probability" that prospect theory problematizes proves to be here something more familiar -- normality of prior distributions. 

In fact, a normal distribution is the only choice of a prior that would annihilate the perceived utility -- as we proceed to prove. 

Imposing Eq.~(\ref{U}) for all $ (q,r)$ enforces $\langle s\rangle_{ax}  = a \langle s\rangle_x$ for all $a \leq 0$. This in turn entails the same equality $\forall a > 0$ as well (just apply it twice with $a_1 =  a_2 =  -\sqrt{a}$). For a one-dimensional argument $x$, this scaling law implies linearity. If the posterior mean has no $x$-derivatives beyond the first, Eq.~(\ref{cumulant_relation}) implies that the posterior distribution has no cumulants beyond the second one, i.e. it is normal. 

Gaussianity of the prior follows immediately through~(\ref{Bayes_rule}).
 
\subsection{First tangent inequality}
\label{sec:first_tangent_inequality}

The derivative of the perceived utility is 

\be
\partial_p U  = \hat{s}(r/p - r) - \hat{s}(-r) - r \hat{s}'(r/p -r) /p
\ee

where $\hat{s}(x) = \langle s \rangle_x$; or in terms of the odds

\be
\label{utility_primed}
\partial_q U  = \frac{r \hat{s}'(qr)+  \hat{s}(-r)}{1 + q} + \frac{- q \hat{s}(-r) - \hat{s}(qr)}{(1+q)^2} 
\ee

Because $U$ must have the opposite sign to $r$ in a left neighborhood of $p=1$ (or right neighborhood of  $q=0$), where it touches zero, it follows that the derivative at that point is of particular importance. From~(\ref{utility_primed}), we see that 
\be 
[\partial_q U]_{q=0} = r \hat{s}'(0) + \hat{s}(-r)
\ee

Now, to comply with the reflection effect this must be negative for gains and positive for losses which finally leads to 

\bea
\label{tangent_inequalities}
\hat{s}(r) &>&  \hat{s}'(0) r \hspace{15pt} \mathrm{if} \ r > 0 
\\
\hat{s}(r) &<& \hat{s}'(0) r \hspace{15pt} \mathrm{if} \ r  < 0 
\eea

I will refer to these inequalities as "tangent inequalities" because they state that the curve must lie above its zero-deviation tangent. 

The tangent inequalities are related to the concave-convex shape described in~\cite{chan2021decision}. Since in that theory $\hat{s}(0)=0$, a second derivative $\hat{s}''(x)$ that has everywhere the same sign as $x$ is a sufficient condition for the forbidden regions to stay untouched. However, it is in general not a necessary condition. The graph of $\hat{s}(x)$ can fluctuate without harm, changing the sign of all its derivatives, as long as it never enters a  forbidden wedge of the Cartesian plane -- located beneath the curve in the upper right quadrant, and above the curve in the lower left quadrant. 

One consequence of this has to do with the behavior at infinity. Namely, if the asymptotic behavior of $\hat{s}(x)$ is linear $\hat{s}(x) =  \beta x + o(x)$, we must have necessarily $  \beta \geq  \hat{s}'(0)$ and, if $\beta = \hat{s}'(0) $, a strictly positive subleading contribution in  the remainder.   

\subsection{Second tangent inequality} 
\label{sec:second_tangent_inequality}

The above inequalities are a lower bound on $\hat{s}(r)/r$. Similarly, we also have an upper bound on $\hat{s}(r)/r$. We will write down the derivation briefly. 
For large q, we can write 

\be U \sim \hat{s}(-r) + \frac{1}{q} \hat{s}(qr) \ee

With a symmetric prior, $\hat{s}(x)$ is antisymmetric, and $U > 0 $ can be written as $ \hat{s}(qr) > - q \hat{s}(-r)  = q \hat{s}(r)$ which in turn can be rearranged (setting $r'=qr$) into 
\be
\hat{s}(r') > \hat{s}(r) + \frac{\hat{s}(r)}{r} (r'-r) 
\ee
or 
\be \frac{\hat{s}(r') - \hat{s}(r)}{r'-r } >  \frac{\hat{s}(r) - \hat{s}(0)}{r - 0 } 
\ee

which  doesn't need to be true for any pair of values $r$ and $r'$, and is only required to hold true in the  limit $ r' \gg r$. 

Now, because we already know that $f$ has to be asymptotically linear, calling its asymptote $\hat{s}(r) \sim \hat{s}'(\infty)  x$, we can write this as 

\be   \hat{s}'(\infty)  > \hat{s}(r) / r \ \ \ \ \forall r \ee 

or, given the continuity of the curve 

\be 
\lim_{r\rightarrow 0 } \hat{s}'(r) = \sup_{\mathbb{R}^+} \frac{ \hat{s}(r) }{r}  
\ee  

\subsection{Restating the  two tangent inequalities} 

\begin{figure}
\centering
\includegraphics[width=.5\linewidth]{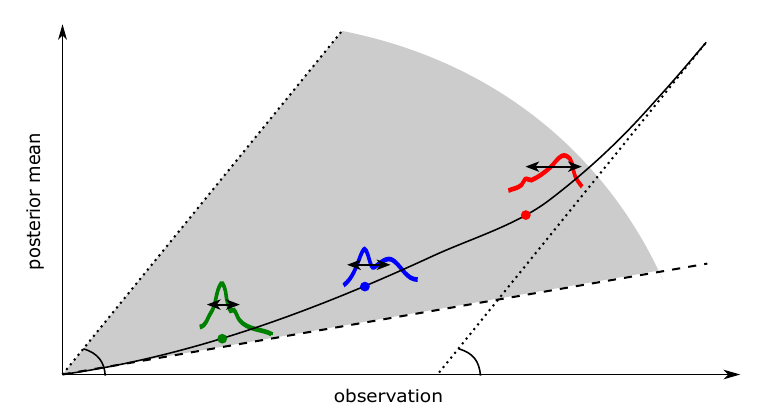}
\caption{
Depiction of the tangent inequalities.}
\label{fig:tangent-inequalities}
\end{figure}

The tangent inequalities, shown graphically in Fig.~\ref{fig:tangent-inequalities}, can be collected in the form 

\be
\label{both_tangent_inequalities}
\boxed{  \hat{s}'(0) < \frac{ \hat{s}(r) }{ r}  <   \hat{s}'(\infty)  } \ee

Obviously, requiring $\hat{s}'(r)$ to be monotonically increasing would be sufficient to ensure this. But it would also be overdemanding. The ratio $\hat{s}(r)/r$  can range widely  over the real axis as long as it never becomes smaller than its zero-deviation value, or larger than the  asymptotic slope.

Notice that, applying~(\ref{cumulant_relation}) with $n=1$, we can rewrite the tangent inequalities as 

\be 
\frac{ \text{Var}  (s | x = 0 ) }{\sigma^2} < 
  \frac{ \langle s \rangle_x }{ x}  
   <  \frac{  \text{Var}  (s | x \rightarrow \infty  )   }{\sigma^2}
\ee

which are direct constraints on the posterior distribution. The next question to tackle is therefore what constraints on the prior are posed by such constraints on the posterior.

\subsection{Second derivative vanishes for zero-deviation outcome}
\label{sec:second_derivative}

The forbidden-region rules derived above teach us something about the derivatives of $\hat{s}(x)$ in the origin. To see this, notice that the inequalities~(\ref{tangent_inequalities}) must apply also for infinitesimally small $|r|$. In other words, we must have

\be
\hat{s}(r) = \hat{s}(0) + \hat{s}'(0) r + \frac{1}{2} \hat{s}''(0) r^2  
+o(r^2)
 \begin{cases} 
> \hat{s}'(0)r \hspace{10pt} \mathrm{if} \ r > 0 \\ 
< \hat{s}'(0)r \hspace{10pt} \mathrm{if} \ r < 0
\end{cases} 
\ee 

Now, we showed in Sec.~\ref{sec:mean_vanishes} that $ \hat{s}(0) = 0$. \textit{Ad  absurdum} suppose $ \hat{s}''(0)$  is different from zero. Then  it becomes the dominant term in these inequalities near the origin, because everything else is zero or cancels out. The tangent inequalities require then the second derivative to be at the same time nonnegative and nonpositive, so it  must vanish. 

The leading term is therefore the one of order $r^3$ or higher. Applying Eq.~(\ref{cumulant_relation}) with $n=2$, notice that 
the vanishing of the second derivative is equivalent to the vanishing of the skewness of the posterior  in the particular case of outcome $r=0$.

\subsection{Open mind theorem (Incompatibility of compact priors)} 
\label{sec:open-mind}

Let us now apply the inequalities (\ref{tangent_inequalities}) to the regime of large values of $|x|$.

Because the variance of likelihood and prior is never zero, $\hat{s}'(0)$  is always positive. In the light of this and of the tangent inequalities (\ref{tangent_inequalities}), considered for large $|r|$, we reach the following conclusion. A prior that leads to a sublinear asymptote for the posterior mean is not admissible because it contradicts experiments. 

Recall now that (as seen in  Sec~\ref{sec:compact-support-and-posterior-mean})   a compact-support prior leads to sublinear asymptotic behavior for the posterior mean. We reach therefore the  conclusion 
( "open-mind theorem") that a compact prior must be ruled out. 

The open mind theorem is depicted in Fig.~\ref{fig:open-mind}. If the prior has compact support as in panel D, the comparative utility function $U$ behaves as in the dashed red curves of panels A and B. 

\subsection{The zero-deviation posterior is leptokurtic} 
\label{sec:suppl_leptokurtic}

Assuming that the fourth central moment of the posterior at $x=0$ doesn't vanish, it must be positive. This is because if $ \hat{s}(0)$  and $ \hat{s}''(0)$  both vanish, we can expand as 
 $ \hat{s}(r) \sim \hat{s}'(0) r  + \frac{1}{6} \hat{s}'''(0) r^3$ and 
and sticking this into the tangent inequalities 
(\ref{tangent_inequalities}) yields
$ \mathrm{sign} (\hat{s}'''(0) r^3 ) = \mathrm{sign}(r) 
$.
 
Applying now Eq.~(\ref{cumulant_relation}) with $n=3$, we can write this result as 

\be 
\mathcal{K} - 3 =  \frac{ \big\langle \big(s - \langle s \rangle \big)^4 \big\rangle }{ \big\langle \big(s - \langle s \rangle \big)^2 \big\rangle^2 } - 3 \ > \ 0
\ee 
 
or more concisely, 
 $ \mathrm{ EK} > 0 $ if EK is the excess kurtosis, i.e. the kurtosis measured from its Gaussian reference value. 

In conclusion, the posterior distribution for zero outcome is leptokurtic i.e. has positive excess kurtosis.

\subsection{Kurtosis of the prior}
 \label{sec:prior_kurtosis} 
  
As explained in~\ref{sec:perceived_utility}, we are pursuing a path inverse to that of the main text, starting from empirically justified requirements on the perceived utility and tracking back what constraints that entails on the prior distribution. In Sec.~\ref{sec:suppl_leptokurtic}, we derived some simple constraints on the moments of the posterior distribution. How do they transfer into constraints upon the prior?

Above, when deducing  that the excess kurtosis of the posterior had to be positive, we did not mention for what value of $\sigma$ we are requiring that positivity. Just like the value of the anchor, so the internally modeled measurement error $\sigma$ is not communicated  to experimenters, and we may only reasonably assume that it fluctuates widely across subjects and contexts. 

Therefore, to take seriously the empirical observations, a constraint derived from them for   the zero-deviation posterior  should be taken as a statement about all the zero-deviation posteriors corresponding to all values of $\sigma$. 

Now call $\vec{R} = (R_1, R_2, \ldots)$ the set of all the cumulants of the prior distribution of $\rho(s)$. Knowing this infinite vector is largely equivalent to knowing the distribution itself. If we call $F_4$ the excess kurtosis of the posterior, now regarded as a function of both $\sigma$ and $\vec{R} $,  the resulting statement about $\vec{R}$ is then

\be 
 F_4(\sigma, \vec{R})> 0 \ \ \ \forall \sigma>0  
\ee 

We thus have to (1)  minimize $F_4$ over $ \sigma$, (2)  require the minimum value to be positive and (3) infer properties of $\vec{R}$ from that.   

If we do that, we can easily  say  that the prior itself must be  leptokurtic ($R_4 >0$). The reason is that in the limit of broad likelihoods ($\sigma \rightarrow \infty$) the posterior is equal to the prior. If the results of experiments have to be valid for large $ \sigma$ as well (i.e. the constraint of the posterior being leptokurtic  must survive) then the prior (which does not depend on $\sigma$) is  leptokurtic. 
 
In conclusion, the prior may not be of the spike-and-slab type but experiments do reveal that, just like the spike-and-slab distribution,  it  has a  positive excess kurtosis (for a  discussion see main text, Sec.~\ref{sec:general}).

\end{document}